\documentclass[11pt,a4paper]{article}\usepackage{jheppub}

\usepackage{amsthm} 	
\usepackage{mathtools} 	
\usepackage{dsfont} 		
\usepackage{slashed}

\def\be#1\ee{\begin{align}#1\end{align}}
\def\pe{\ \:\phantom{=}}
\def\ba{\begin{eqnarray}}
\def\ea{\end{eqnarray}}
\def\la{\langle}
\def\ra{\rangle}
\def\teps{\tilde{\varepsilon}}

\def\f{\frac}
\def\lb{\big\lbrace}
\def\rb{\big\rbrace}

\def\openone{\mathds{1}}

\def\nn{\nonumber}
\def\q{\qquad}

\def\ip{\lrcorner\,}

\def\gL{\mathcal{L}^\text{g}}

\def\tr{\mathrm{tr}}

\def\de{\mathrm{d}}
\def\De{\mathrm{D}}
\def\SU{\mathrm{SU}}

\def\su{\mathfrak{su}}
\def\so{\mathfrak{so}}

\def\E{\mathcal{E}}
\def\F{\mathcal{F}}
\def\G{\mathcal{G}}
\def\H{\mathcal{H}}

\def\L{\mathcal{L}}
\def\O{\mathcal{O}}
\def\Q{\mathcal{Q}}

\def\X{\mathcal{X}}
\def\Y{\mathcal{Y}}

\renewcommand{\figurename}{Figure}

\begin{document}

\title{Edge modes and corner ambiguities in\\ 3d Chern--Simons theory and gravity}

\author{Marc Geiller}
\affiliation{Perimeter Institute for Theoretical Physics,\\ 31 Caroline Street North, Waterloo, Ontario, Canada N2L 2Y5}
\emailAdd{mgeiller@perimeterinstitute.ca}

\abstract{Boundaries in gauge field theories are known to be the locus of a wealth of interesting phenomena, as illustrated for example by the holographic principle or by the AdS/CFT and bulk-boundary correspondences. In particular, it has been acknowledged for quite some time that boundaries can break gauge invariance, and thereby turn gauge degrees of freedom into physical ones. There is however no known systematic way of identifying these degrees of freedom and possible associated boundary observables. Following recent work by Donnelly and Freidel, we show that this can be achieved by extending the covariant Hamiltonian formalism so as to make it gauge-invariant under arbitrary large gauge transformations. This can be done at the expense of extending the phase space by introducing new boundary fields, which in turn determine new boundary symmetries and observables. We present the general framework behind this construction, and find the conditions under which it can be applied to an arbitrary Lagrangian. By studying the examples of Abelian Chern--Simons theory and first order three-dimensional gravity, we then show that the new boundary observables satisfy the known corresponding Kac--Moody affine algebras. This shows that this new extended phase space formulation does indeed properly describe the dynamical boundary degrees of freedom, and gives credit to the results which have been previously derived in the case of diffeomorphism symmetry. We expect that this systematic understanding of the boundary symmetries will play a major role for the quantization of gravity in finite regions.}

\maketitle

\addtocontents{toc}{\protect\setcounter{tocdepth}{2}} 

\section{Introduction}

The status of boundaries in modern theoretical physics has evolved from being simply ``the place where we set boundary conditions" to the locus of a wealth of phenomena whose richness and physical relevance is becoming increasingly apparent. This is epitomized for example in the holographic principle \cite{Susskind:1994vu,Bousso:2002ju}, the AdS/CFT correspondence \cite{Brown:1986nw,Coussaert:1995zp,Maldacena:2003nj}, the bulk-boundary correspondence of condensed matter \cite{Wen:1992vi,Wen:2004ym,2017arXiv170200673K}, or the study of (entanglement) entropy \cite{Srednicki:1993im,Eisert:2008ur,Solodukhin:2011gn,Donnelly:2011hn,Donnelly:2014gva,Delcamp:2016eya}. Depending on the system, the boundaries of interest can be inner to the spacetime (and provide for example a quasi-local description of black holes \cite{Booth:2005qc}), at infinity (and describe the asymptotic geometry of spacetime \cite{Ashtekar:1978zz,Ashtekar:1991hf,Ashtekar:1991vb,Ashtekar:2014zsa}), or at finite distances (and delimitate subsystems). To this list one can also add boundaries in the form of defects of arbitrary co-dimension, which can support local excitations in topological quantum field theories \cite{Atiyah:1989vu,Schwarz:2000ct,Segal:2001aa,SchommerPries:2011np} and play now a central role in condensed matter \cite{Walker:2011mda,Barkeshli:2014cna,Wen:2016ddy} and quantum gravity \cite{Dittrich:2016typ,Delcamp:2016lux,Delcamp:2016yix,Dittrich:2017nmq}. Although there is no known framework to describe at once the new physics that can emerge on boundaries, one unifying thread has to do with the notion of gauge.

Heuristically, the role of gauge transformations is to identify field configurations which would have otherwise been deemed inequivalent. However, when defining gauge field theories on manifolds with boundaries these latter can break gauge invariance and thereby turn certain gauge degrees of freedom into physical ones. These are for example the famous conformal edge currents of Chern--Simons theory \cite{Wen:1992vi,Balachandran:1991dw}. In the context of gravity, these ``would-be-gauge'' degrees of freedom have been put forward as candidates to explain the origin of black hole entropy \cite{Carlip:1994gy,Balachandran:1994up,Balachandran:1995qa}. By using the Chern--Simons formulation of three-dimensional gravity \cite{MR974271,Achucarro:1987vz} Carlip was able to construct an explicit realization of this idea \cite{Carlip:1994gy}. His construction relies partly on the fact that the Chern--Simons action is not gauge-invariant and gives rise on the boundary to a Wess--Zumino--Novikov--Witten (WZNW) theory \cite{Moore:1988uz,MR1025431,Gawedzki:2001ye,Arcioni:2002vv}. A more general result is actually known to hold. In a seminal paper, Brown and Henneaux proved that the asymptotic symmetries of three-dimensional AdS spacetime were described by two copies of a Virasoro algebra \cite{Brown:1986nw}, and it was shown later on that the corresponding boundary dynamics is given by a Liouville conformal field theory \cite{Coussaert:1995zp,Banados:1998ta}. This constitutes the first example of a realization of AdS/CFT. These examples from three-dimensional gravity are all the more surprising because they give rise to infinitely-many degrees of freedom on the boundary while the bulk can only have finitely-many owing to the topological nature of the theory. Similar proposals exist for the description of four-dimensional black hole entropy \cite{Carlip:2011vr}, but are more complicated due to the unavailability of a Chern--Simons formulation and the need to work with diffeomorphisms \cite{Carlip:2017xne,Carlip:2016lnw}.

It has been known for a while that by slightly generalizing Noether's two theorems \cite{Noether:1918zz} it is possible to consistently assign conserved charges to local gauge symmetries \cite{Iyer:1994ys,Barnich:2001jy}. Recently, a huge momentum was gained following the realization that a carefully analysis of boundary conditions and large gauge transformations (i.e. gauge transformations which do not vanish on the boundary) leads to new conserved charges in an unexpectedly broad variety of theories (see \cite{Avery:2015rga} and references therein), including most notably QED \cite{Barnich:2001jy,Lysov:2014csa,Kapec:2015ena} and gravity \cite{Hawking:2016msc}. These results build up on the existence in these theories of infinite-dimensional asymptotic symmetry algebras, like the Bondi--Metzner--Sachs (BMS) algebra in the asymptotically flat gravitational case \cite{Bondi:1962px,Sachs:1962zza,Sachs:1962wk}. The existence of well-defined conserved charges associated with certain residual gauge transformations means that these latter are actually best thought of as symmetries, which therefore map between physically inequivalent field configurations. Understanding the origin and the physical implications of these forgotten degrees of freedom, or ``soft hairs'' as they are now known, is of primordial importance. As far as we are aware however, there exists so far no systematic and universal understanding of the nature of the conserved charges and asymptotic symmetries which should be considered as physical given an arbitrary theory. Rather, the conserved charges which have been constructed so far always involve some extra structure, in the sense that they live on specific regions of spacetime and require specific boundary conditions (see for example \cite{Strominger:2017zoo}).

In the present paper, we will focus on yet another type of boundary charges, namely those defined at finite distances. Interestingly, the crucial role played by such boundary charge degrees of freedom has been impressively illustrated in numerous (lattice) gauge theory computations of entanglement entropy \cite{Buividovich:2008gq,Donnelly:2011hn,Lewkowycz:2013laa,Donnelly:2014fua,Donnelly:2014gva,Donnelly:2015hxa,Soni:2015yga,Aoki:2015bsa,Ghosh:2015iwa}. In short, if the contribution of boundary degrees of freedom is forgotten one ends up undercounting the entropy. This is remedied by considering so-called extended Hilbert spaces which contain information about gauge transformations with non-vanishing support on the boundary. While most of the results in these approaches are concerned with the Hilbert spaces (attached to local regions or subsystems) of quantum lattice gauge theories, Donnelly and Freidel have recently proposed a classical and continuum analysis of the mechanism at play, and applied it to gravity in metric variables \cite{Donnelly:2016auv}. They have in particular defined an extended phase space containing new boundary degrees of freedom, derived new boundary observables for Yang--Mills theory and gravity, and found in this latter case that they are described by an unexpectedly large symmetry group. This can potentially have very important consequences for the quantization of gravity in finite regions, and therefore deserves a very thorough analysis. In the present work we would like to give some more flesh to their argument, and to compare its consequences with previously known results about boundary observables in gauge theory and gravity. We therefore ask the following two questions:

\textit{i) Starting from the Lagrangian of any gauge theory, how to construct in a definite manner an extended phase space containing relevant boundary degrees of freedom?}

\textit{ii) What is the interpretation of the new boundary observables and symmetries which appear in this extended phase space?}

The first question was already studied in \cite{Donnelly:2016auv} on the particular examples of Yang--Mills theory and second order gravity. Here we would like to develop a general understanding of this construction without focusing on a particular example. This will force us to think carefully about boundary terms, corner ambiguities, possible gauge-non-invariance of the Lagrangian, and the definition of the conserved pre-symplectic form. This will be the first part of our work.

The second question on the other hand has to do with specific examples. As mentioned earlier, there are many theories for which the boundary observables and degrees of freedom are (believed to be) known. This includes for example the edge modes of Abelian Chern--Simons theory \cite{Wen:1992vi,Balachandran:1991dw,Tong:2016kpv} and the WZNW gauge fields of non-Abelian Chern--Simons theory \cite{Carlip:1994gy,Arcioni:2002vv}. In gravity, observables were constructed in \cite{Balachandran:1994up,Balachandran:1995qa} with metric variables and in \cite{Husain:1997fm} with first order connection and triad variables. This thus begs the question of the relationship between these ``old'' observables and the ``new'' observables of \cite{Donnelly:2016auv}. The second part of our work will therefore be devoted to the study of this question in Chern--Simons theory and first order three-dimensional gravity. We will find that the new observables coming from the extended phase space are a ``dressed'' version of the previously-known observables, but that they satisfy the same current algebra. The case of non-Abelian Chern--Simons theory can be treated along the same ways, and we outline the main steps of the construction in appendix \ref{appendix:CS}. We will also see that the extended phase space obtained by introducing additional boundary degrees of freedom enables to obtain conceptual clarity as to the role of gauge transformations versus that of gauge symmetries.

In other words, while it has been known for quite some time how to describe the boundary dynamics of Chern--Simons theory (i.e. the WZNW action) and how to describe Hamiltonian boundary observables in Chern--Simons theory and gravity, here we will take these steps further and introduce a new set of boundary variables, which are found through the requirement of gauge-invariance of the so-called covariant Hamiltonian framework, and which will lead to a ``dressing'' of the previously-known boundary observables. As we will show, for the construction of this extended phase space containing the new dressing boundary variables, it is not enough to simply consider the gauge-invariant Lagrangian of the theory. Instead, let us now present how we will proceed:

In order to simplify the understanding of the interplay between gauge transformations and boundaries, we will focus (as in \cite{Donnelly:2016auv}) on $(d-2)$-dimensional\footnote{We denote by $d$ the dimension of spacetime.} spatial boundaries at finite distances. This saves us from having to discuss issues of convergence for e.g. the symplectic structure, which would furthermore require to choose a particular asymptotic spacetime structure (e.g. asymptotically flat or AdS).

The fate of gauge transformations is best studied in the covariant Hamiltonian (or covariant phase space) formalism of \cite{Crnkovic:1987tz,Iyer:1994ys,Barnich:2001jy}. In this elegant framework, a central role is played by the so-called pre-symplectic potential. This is a ($d-1$)-form in spacetime and a 1-form in field-space which is determined by the Lagrangian of the theory, and in turn enables to derive its Noether currents, its symplectic structure, and the generators of infinitesimal gauge transformations. We will therefore start our work in section \ref{sec:2} with a review of the covariant Hamiltonian formalism, and devote particular attention to boundary (i.e. $(d-1)$-dimensional) and corner (i.e. $(d-2)$-dimensional) terms in the Lagrangian. This is usually overlooked in most treatments, but will enable us to identify possible ambiguities or subtleties in e.g. theories whose Lagrangian is not gauge-invariant. Following the insight of \cite{Donnelly:2016auv}, we will show that important information about gauge transformations on the boundary can be obtained by considering finite and field-dependent gauge transformations\footnote{The fact that the construction of \cite{Donnelly:2016auv} could more elegantly and rigorously be understood in terms of field-dependent gauge transformations was actually pointed out and studied in \cite{Gomes:2016mwl}.} of the pre-symplectic potential. This means that we will allow for non-vanishing field-space variations $\delta$ of the finite gauge parameters, lurking towards the idea that they could be promoted to the status of degrees of freedom and thereby contribute to the symplectic structure. The precise way in which this should be done can be understood by noticing that the pre-symplectic potential is not gauge-invariant under such finite field-dependent transformations. Amongst our new results we will in particular derive the general form of this transformation, and show that it is possible to introduce boundary degrees of freedom that ``extend'' the pre-symplectic potential and ensure its gauge invariance. We will then analyse the conditions under which this extended pre-symplectic potential can be used to assign vanishing Noether charges\footnote{The fact that gauge transformations come with vanishing Noether charges in this extended framework has also been pointed out in the construction of \cite{Gomes:2016mwl}, which however obtains gauge-invariance by using a covariant derivative on field-space. As we will see, a more precise statement is actually that the charges are at least gauge-invariant, and at best vanishing.} to gauge transformations and to construct a conserved pre-symplectic form. As we will see, this will amount to relaxing the usual conservation criterion for the pre-symplectic form, and the new boundary degrees of freedom will act as compensating fields for the symplectic flux which is leaked through the boundary. This is precisely realizing the idea that boundary conditions should be relaxed on the boundary, and additional fields introduced in order to keep track of the information flowing through the boundary from one subsystem to the neighboring one.

In section \ref{sec:3}, we apply the general results of section \ref{sec:2} to Abelian Chern--Simons theory. This constitues an ideal testbed because there exists already a known Lagrangian \cite{Carlip:1994gy,Tong:2016kpv} and Hamiltonian \cite{Balachandran:1991dw,Balachandran:1995qa} description of the boundary degrees of freedom and of their dynamics. In particular, these descriptions rely respectively on the gauge-non-invariance of the Lagrangian, and on the Regge--Teitelboim criterion for functional differentiability of the Hamiltonian generators \cite{Regge:1974zd}. We will first review how these arguments arise, and then show that the extended phase space of \cite{Donnelly:2016auv} and section \ref{sec:2} gives a much more systematic construction, which in particular does not require a choice of boundary conditions or restrictions on the gauge parameters. Out of this construction, we will obtain the dressed boundary observables and their algebra, which is nothing but the affine Kac--Moody algebra of the gauge group \cite{Walton:1999xc}. We therefore recover known results, but with the following conceptual advantage: the gauge transformations are now generated by vanishing generators with a closed algebra and assigned gauge-invariant (or vanishing) Noether charges, and the new boundary degrees of freedom of the extended phase space give rise to a new boundary symmetry. The generators of this latter are precisely the boundary observables, which satisfy the current algebra.

Three-dimensional gravity in its first order formulation is studied in section \ref{sec:4}. As is well-known, this is a topological field theory \cite{Horowitz:1989ng,MR1637718}, which as such is invariant under three types of (non-independent) transformations: diffeomorphisms, $\SU(2)$ (in the Euclidean case) gauge transformations, and translations. We start by recalling the infinitesimal and finite form of these gauge transformations, and give for the first time the expression for the finite translations in the case of a non-vanishing cosmological constant (the infinitesimal version appears of course in \cite{MR974271}). After repeating the usual Hamiltonian arguments which lead to boundary observables, we proceed following section \ref{sec:2} with the construction of the gauge-invariant extended pre-symplectic potential. This is done for two choices of independent gauge transformations: first for $\SU(2)$ gauge transformations and translations, and then for $\SU(2)$ gauge transformations and diffeomorphisms. We then study the new observables which appear on the extended phase space. For simplicity we focus on $\SU(2)$ gauge transformations and translations, and leave the study of diffeomorphisms (and therefore comparison with \cite{Donnelly:2016auv}) for future work. The result is that the new observables are precisely a dressed (i.e. gauge transformed by the new boundary fields) version of the observables appearing in \cite{Husain:1997fm}, to which they reduce if the new boundary degrees of freedom are ignored and trivialized. Their algebra is furthermore given by the affine Kac--Moody algebra of $\text{ISU(2)}$ (in the case of a vanishing cosmological constant).

While the affine Kac--Moody algebras which we obtain in this work as boundary symmetry algebras are already known to appear from the Regge--Teitelboim condition of functional differentiability of the constraints, what we show here is that the boundary fields of the extended phase space allow for a new realization of these algebras in terms of new boundary observables. What will therefore be important for future work is to understand the physical role played by these boundary fields and observables.

We will present our conclusions and future research directions in section \ref{sec:5}. The appendices contain some useful formulas as well as the proof of some more lengthy calculations, the main formulas needed in order to extend this construction to non-Abelian Chern--Simons theory, and a brief treatment of diffeomorphisms.

\section{Covariant Hamiltonian formalism and corner ambiguities}
\label{sec:2}

In this first section, we present the general framework underlying our construction and that of \cite{Donnelly:2016auv}. This relies on well-established results in the so-called covariant Hamiltonian formalism (also known as the covariant phase space formalism), which we here review while carefully keeping track of boundary and corner terms. In particular, we will see that a specific corner ambiguity can be fixed by demanding that the covariant Hamiltonian framework be ``preserved'' by \textit{finite field-dependent} gauge transformations, and that this can be done at the expense of introducing new boundary fields.

The reader is free to skip this formal presentation of the results and jump straight to the example of Abelian Chern--Simons theory in section \ref{sec:3}.

\subsection{Conserved pre-symplectic form}

The covariant Hamiltonian formalism is a way of studying the generators of infinitesimal gauge transformations, along with their charges and their algebra, without the need to resort to a non-manifestly-covariant decomposition between space and time. A central object in this formalism is the conserved pre-symplectic form, whose construction we now describe.

For the sake of generality, let us consider a Lagrangian $L[\Phi]$ depending on a set of fields $\Phi$, and add to it a possible boundary term $b[\Phi]$. The variation of the Lagrangian with its boundary term is then given by\footnote{Throughout this work, we will always keep track with subscripts of the terms and ambiguities which contribute to the right-hand side of the various equalities and definitions. A subscript $b$ denotes for example the presence of a contribution from a boundary term, a subscript $\ell$ will denote the contribution from a boundary Lagrangian with new fields, a subscript $c$ will denote the contribution coming from a corner ambiguity, etc. Since these objects will all be clearly defined, there should be no ambiguity as to what the subscripts refer to. Furthermore, while section \ref{sec:2} is slightly more formal with these notations, sections \ref{sec:3} and \ref{sec:4} will provide all the concrete examples.}
\be\label{variation of Lagrangian}
\delta L_b[\Phi]=\delta(L[\Phi]+\de b[\Phi])=E[\Phi]\wedge\delta\Phi+\de\theta_{b,c}[\Phi,\delta\Phi].
\ee
On the right-hand side, the first term identifies the equations of motion $E[\Phi]$, and the second term identifies the pre-symplectic potential\footnote{From now on we will often refer to this object as ``the potential''.}
\be\label{definition general potential}
\theta_{b,c}[\Phi,\delta\Phi]\coloneqq\theta_b[\Phi,\delta\Phi]+\de c[\Phi,\delta\Phi]\coloneqq\theta[\Phi,\delta\Phi]+\delta b[\Phi]+\de c[\Phi,\delta\Phi].
\ee
This object is a $(d-1,1)$-form, i.e. a $(d-1)$- form in spacetime and a 1-form in field space. Since it appears in \eqref{variation of Lagrangian} via its exterior derivative, its identification is ambiguous up to the addition of a closed form, which is the so-called corner ambiguity.

From the knowledge of a potential, we can then construct the associated pre-symplectic current $(d-1,2)$-form\footnote{We use here the shorthand notation $\delta P\delta Q=\delta_1P\delta_2Q-\delta_2P\delta_1Q$ for the 2-forms in field space. Furthermore, since in the 2-forms in field space all the fields are acted on by a variation $\delta$, we use the notation $\omega[\Phi,\delta_1,\delta_2]$ instead of $\omega[\delta_1\Phi,\delta_2\Phi]$, which will eventually prove to be lighter.} by taking a field space derivative, i.e.
\be\label{pre-symplectic current}
\omega_{b,c}[\Phi,\delta_1,\delta_2]\coloneqq\delta\theta_{b,c}[\Phi,\delta\Phi].
\ee
From \eqref{definition general potential} and the fact that $\delta^2=0$, one can see that this reduces to
\be
\omega_{b,c}[\Phi,\delta_1,\delta_2]=\delta\theta[\Phi,\delta\Phi]+\de\delta c[\Phi,\delta\Phi]=\omega[\Phi,\delta_1,\delta_2]+\de\delta c[\Phi,\delta\Phi]=\omega_c[\Phi,\delta_1,\delta_2].
\ee
An important property is that the pre-symplectic current is closed on-shell. More precisely, we have
\be
\de\omega_{b,c}[\Phi,\delta_1,\delta_2]=\de\omega_c[\Phi,\delta_1,\delta_2]=\de\omega[\Phi,\delta_1,\delta_2]\simeq0.
\ee
Notice that these three equalities have a different origin. The first one is a strict equality, and comes from the fact that the boundary term has completely dropped from the pre-symplectic current. The second one comes from the fact that the corner ambiguity is the addition of a total exterior derivative. And finally, the third one is the statement that the pre-symplectic current is closed on-shell.

Let us now consider a $d$-dimensional region $N\subset M$ of spacetime bounded by $\partial N=\Sigma_1\cup\Sigma_2\cup B$, where $\Sigma_1$ and $\Sigma_2$ are two $(d-1)$-dimensional space-like hypersurfaces, and $B$ is a boundary (which can also consist of an inner boundary and/or a time-like boundary at infinity). Integrating the $(d,2)$-form $\de\omega[\Phi,\delta_1,\delta_2]$ over $N$, we get that
\be\label{integration of omega}
\int_N\de\omega[\Phi,\delta_1,\delta_2]=\int_{\partial N}\omega[\Phi,\delta_1,\delta_2]=\left(\int_{\Sigma_2}-\int_{\Sigma_1}+\int_B\right)\omega[\Phi,\delta_1,\delta_2]\simeq0.
\ee
At this point, one has to impose boundary conditions in order to deal with the contribution at $B$. Note that these boundary conditions do not know about the boundary conditions on the fields at $\partial M$ which guarantee that the variational principle is well-defined, and in particular do not know about the existence of a possible boundary term $\de b[\Phi]$ in the Lagrangian. One possibility is to assume that the pre-symplectic flux coming from $B$ is vanishing. In this case, we get that the pre-symplectic $(0,2)$-form, defined as\footnote{Alternatively one can first define the $(0,1)$-form
\be\label{pre-symplectic 1-form}
\Theta_\Sigma[\Phi,\delta\Phi]\coloneqq\int_\Sigma\theta[\Phi,\delta\Phi],
\ee
and, assuming that the specification of $\Sigma$ is field-independent so that the field-space variation $\delta$ can freely pass through the integral, we get that $\Omega_\Sigma[\Phi,\delta_1,\delta_2]=\delta\Theta_\Sigma[\Phi,\delta\Phi]$. As pointed out and discussed in \cite{Donnelly:2016auv,Gomes:2016mwl}, extra care has to be taken in the case where $\Sigma$ is defined in a relational manner through the field content of the theory.}
\be\label{bare conserved form}
\Omega_\Sigma[\Phi,\delta_1,\delta_2]\coloneqq\int_\Sigma\omega[\Phi,\delta_1,\delta_2],
\ee
does not depend on the hypersurface $\Sigma$ when the equations of motion are satisfied. This is the conserved pre-symplectic form.

One weaker possibility is to have boundary conditions at $B$ which are such that $\omega[\Phi,\delta_1,\delta_2]\big|_B=\de w[\Phi,\delta_1,\delta_2]$. In this case, we get that
\be
\int_B\omega[\Phi,\delta_1,\delta_2]=\int_{\partial B}w[\Phi,\delta_1,\delta_2]=\left(\int_{S_1}-\int_{S_2}\right)w[\Phi,\delta_1,\delta_2]
\ee
where $S_i=\Sigma_i\cap B$, and the conserved pre-symplectic form is given by
\be\label{conserved form with corner}
\Omega_{\Sigma,S}[\Phi,\delta_1,\delta_2]\coloneqq\int_\Sigma\omega[\Phi,\delta_1,\delta_2]-\int_Sw[\Phi,\delta_1,\delta_2].
\ee
This happens for example in the case of isolated horizons \cite{Corichi:2016zac,Corichi:2013zza}.

In what follows, we will start by assuming that we have a conserved pre-symplectic form as defined in \eqref{bare conserved form}, but this construction will eventually be relaxed when considering the corner ambiguity. We can now explain how to obtain the generators of gauge transformations.

\subsection{Charges and generators of infinitesimal transformations}

Let $\epsilon$ denote the parameter of an infinitesimal gauge transformation\footnote{Unless otherwise stated, the infinitesimal gauge transformations will always be \textit{field-independent} in this work. We will however be led later on to consider finite \textit{field-dependent} transformations in order to achieve our construction and to illustrate some subtleties which are often overlooked in the literature.} acting on the fields as $\delta_\epsilon\Phi$. In the case of diffeomorphisms, $\epsilon$ will be a vector field $\xi$, while in the case of internal gauge transformations it will be a Lie algebra element $\alpha$. The generator of this infinitesimal gauge transformation is a function $\H[\epsilon]$ over the phase space, whose Poisson bracket with the fields is
\be
\lb\H[\epsilon],\Phi\rb=\delta_\epsilon\Phi.
\ee
A fundamental result of the covariant Hamiltonian framework is that this generator is defined by the variational formula
\be\label{gauge generator and 2-form}
\slashed{\delta}\H[\epsilon]=\Omega_\Sigma[\Phi,\delta,\delta_\epsilon].
\ee
From this, one can see that the algebra of these generators is given by
\be\label{gauge generator algebra}
\lb\H[\epsilon_1],\H[\epsilon_2]\rb=\delta_{\epsilon_1}\H[\epsilon_2]=\Omega_\Sigma[\Phi,\delta_{\epsilon_1},\delta_{\epsilon_2}].
\ee
In appendix \ref{appendix:generators} we elaborate a bit on these equalities, and in particular generalize them to the case of a pre-symplectic form containing a boundary piece. This will be important for our discussion in the next section. Also, note that when contracting the symplectic $(0,2)$-form $\Omega_\Sigma$ with a variational vector $\delta_\epsilon$, the quantity $\slashed{\delta}\H[\epsilon]$ which is obtained is not guaranteed to be integrable, i.e. to be a total variation $\delta$ of some expression $\H[\epsilon]$. This requires in general additional integrability conditions. This is the reason for which we use the notation $\slashed{\delta}$.

Another important result is that \eqref{gauge generator and 2-form} is always given by the sum of a $(d-1)$-dimensional (spatial) bulk integral involving the equations of motion and a $(d-2)$-dimensional boundary (or corner) integral \cite{Iyer:1994ys,Barnich:2001jy,Barnich:2007bf,Compere:2009dp}. This boundary integral, which is the on-shell value of the generator, is the charge of the transformation. We would now like to recall the proof of this result. This will be the occasion of introducing the Noether charges, and of seeing how boundary terms and corner ambiguities propagate throughout the calculations.

Consider an infinitesimal internal gauge transformation parametrized by $\alpha$ and acting on the Lagrangian by producing a total derivative, i.e.
\be
\delta_\alpha L[\Phi]=\de m[\Phi,\alpha].
\ee
For diffeomorphisms parametrized by $\xi$, the infinitesimal action is given by the Lie derivative, i.e. $\delta_\xi=\L_\xi=\de(\xi\ip\cdot)+\xi\ip(\de\cdot)$, so we have
\be
\delta_\xi L[\Phi]=\de(\xi\ip L[\Phi]).
\ee
In order to treat both cases at once, we will consider a parameter $\epsilon$ and the gauge transformation $\delta_\epsilon=\delta_\alpha+\delta_\xi$. When acting on the Lagrangian with a boundary term, this becomes
\be
\delta_\epsilon L_b[\Phi]=\delta_\epsilon(L[\Phi]+\de b[\Phi]).
\ee
For internal gauge transformations we have $\delta_\alpha(\de b[\Phi])=\de(\delta_\alpha b[\Phi])$, while for diffeomorphisms $\delta_\xi(\de b[\Phi])=\de\big(\xi\ip(\de b[\Phi])\big)$. This leads to
\be
\delta_\epsilon L_b[\Phi]=\de\big(\xi\ip L[\Phi]+\xi\ip(\de b[\Phi])+m[\Phi,\alpha]+\delta_\alpha b[\Phi]\big).
\ee
On the other hand, using $\de(\delta_\xi b[\Phi])=\de\big(\de(\xi\ip b[\Phi])+\xi\ip(\de b[\Phi])\big)$ in \eqref{variation of Lagrangian} leads to
\be\label{arbitrary diff and gauge variation Lagrangian}
\delta_\epsilon L_b[\Phi]=E[\Phi]\wedge\delta_\epsilon\Phi+\de\big(\theta[\Phi,\delta_\epsilon\Phi]+\de(\xi\ip b[\Phi])+\xi\ip(\de b[\Phi])+\delta_\alpha b[\Phi]\big).
\ee
Putting these two equations together finally leads to
\be
E[\Phi]\wedge\delta_\epsilon\Phi+\de\big(\theta_c[\Phi,\delta_\epsilon\Phi]+\de(\xi\ip b[\Phi])-\xi\ip L[\Phi]-m[\Phi,\alpha]\big)=0.
\ee
We can now identify the Noether current $(d-1,0)$-form associated with the transformation $\delta_\epsilon$. It is defined as
\be\label{Noether current}
J_{b,c}[\Phi,\epsilon]\coloneqq\theta_c[\Phi,\delta_\epsilon\Phi]+\de(\xi\ip b[\Phi])-\xi\ip L[\Phi]-m[\Phi,\alpha],
\ee
and we can see that on-shell we have
\be
\de J_{b,c}[\Phi,\epsilon]\simeq0.
\ee
This in turn implies that we can write
\be\label{J and Q on-shell}
J_{b,c}[\Phi,\epsilon]\simeq\de Q_{b,c}[\Phi,\epsilon],
\ee
where the $(d-2,0)$-form
\be
Q_{b,c}[\Phi,\epsilon]\coloneqq Q[\Phi,\epsilon]+\xi\ip b[\Phi]+c[\Phi,\delta_\epsilon\Phi]
\ee
is the Noether charge density associated to $\epsilon$. From this expression, one can see that the Noether charge density is only sensitive to boundary terms in the case of diffeomorphisms. Notice however that the corresponding contribution is only present because we have kept the corner term in \eqref{arbitrary diff and gauge variation Lagrangian}, and represents therefore an ambiguity \cite{Iyer:1994ys}. These ingredients are all we need in order to compute the Noether charge associated with a transformation $\delta_\epsilon$. However, in order to prove the result concerning the form of $\delta\H[\epsilon]$, we need to go through a few more equations.

First, we would like to have an equality which defines the off-shell part of \eqref{J and Q on-shell}. For this, we use the fact that we can always rewrite the term involving the equations of motion and the gauge transformation of the fields as
\be\label{setup for Noether's second}
E[\Phi]\wedge\delta_\epsilon\Phi=\epsilon N[E,\Phi]-\de P[E,\Phi,\epsilon].
\ee
Noether's second theorem guarantees that $N[E,\Phi]=0$, which implies in turn that we can write
\be\label{Noether in terms of Q and P}
J_{b,c}[\Phi,\epsilon]=P[E,\Phi,\epsilon]+\de Q_{b,c}[\Phi,\epsilon].
\ee
Now, let us consider the equality
\be
0
&=\delta\delta_\alpha L_b[\Phi]-\delta_\alpha\delta L_b[\Phi]\nn\\
&=\de(\delta m[\Phi,\alpha]+\delta\delta_\alpha b[\Phi]-\delta_\alpha\theta_c[\Phi,\delta\Phi]-\delta_\alpha\delta b[\Phi])\nn\\
&=\de(\delta m[\Phi,\alpha]-\delta_\alpha\theta_c[\Phi,\delta\Phi]).
\ee
This implies that there exists a $(d-2,1)$-form $M_c[\Phi,\alpha,\delta\Phi]$ such that
\be\label{definition closed M}
\delta m[\Phi,\alpha]-\delta_\alpha\theta_c[\Phi,\delta\Phi]\eqqcolon\de M_c[\Phi,\alpha,\delta\Phi].
\ee
Now, the variation of the Noether current is given by
\be
\delta J_{b,c}[\Phi,\epsilon]=\delta\theta_c[\Phi,\delta_\epsilon\Phi]+\delta\de(\xi\ip b[\Phi])-\delta(\xi\ip L[\Phi])-\delta m[\Phi,\alpha].
\ee
The third term can be rewritten using
\be
\delta(\xi\ip L[\Phi])
&=\xi\ip\delta L[\Phi]\nn\\
&=\xi\ip(E[\Phi]\wedge\delta\Phi)+\xi\ip(\de\theta_c[\Phi,\delta\Phi])\nn\\
&=\xi\ip(E[\Phi]\wedge\delta\Phi)+\L_\xi\theta_c[\Phi,\delta\Phi]-\de(\xi\ip\theta_c[\Phi,\delta\Phi])\nn\\
&=\xi\ip(E[\Phi]\wedge\delta\Phi)+\delta_\xi\theta_c[\Phi,\delta\Phi]-\de(\xi\ip\theta_c[\Phi,\delta\Phi]),
\ee
and the fourth term can be rewritten using \eqref{definition closed M}. One can see here that we have explicitly used the field-independence $\delta\xi=0$. Putting this together leads to
\be
\delta J_{b,c}[\Phi,\epsilon]=\omega_c[\Phi,\delta,\delta_\epsilon]-\xi\ip(E[\Phi]\wedge\delta\Phi)+\de(\xi\ip\delta b[\Phi]+\xi\ip\theta_c[\Phi,\delta\Phi]-M_c[\Phi,\alpha,\delta\Phi]).
\ee
Finally, rearranging the terms in this equation and using the variation of \eqref{Noether in terms of Q and P} leads to
\be
\omega_c[\Phi,\delta,\delta_\epsilon]
&=\delta P[E,\Phi,\epsilon]+\xi\ip(E[\Phi]\wedge\delta\Phi)+\de(\delta Q_{b,c}[\Phi,\epsilon]-\xi\ip\delta b[\Phi]-\xi\ip\theta_c[\Phi,\delta\Phi]+M_c[\Phi,\alpha,\delta\Phi])\nn\\
&\simeq\de(\delta Q_{b,c}[\Phi,\epsilon]-\xi\ip\delta b[\Phi]-\xi\ip\theta_c[\Phi,\delta\Phi]+M_c[\Phi,\alpha,\delta\Phi]),
\ee
where we have used \textit{both} the equations of motion $E[\Phi]\simeq0$ and the linearized equations of motion\footnote{While the generator $\delta H[\Phi,\epsilon]$ has to be defined for arbitrary variations, i.e. without using the linearized equations of motion, the corresponding charge is defined by an integration in the solution subspace of field-space, and can therefore be defined using the linearized equations of motion.} $\delta P[E,\Phi,\epsilon]\simeq0$. This shows the desired result, namely that
\be\label{charge as boundary integral}
\slashed{\delta}\H[\Phi,\epsilon]=\Omega_\Sigma[\Phi,\delta,\delta_\epsilon]\simeq\int_{\partial\Sigma}(\delta Q_b[\Phi,\epsilon]-\xi\ip\delta b[\Phi]-\xi\ip\theta[\Phi,\delta\Phi]+M[\Phi,\alpha,\delta\Phi]).
\ee
Notice that here and in \eqref{gauge generator and 2-form} we have defined the variation of the generator in terms of the pre-symplectic form without a corner ambiguity. The goal of this work is precisely to determine which corner ambiguity should be used in the definition of the pre-symplectic form, and how this affects the generators and their charges.

Furthermore, notice that the above equation is actually not so useful for computing the charges, since for this one can simply compute $\Omega_\Sigma[\Phi,\delta,\delta_\epsilon]$ and then identify the boundary integral. However, it has the advantage of showing the general result that charges associated with gauge transformations are given by boundary integrals, and that, when considering corner ambiguities in the pre-symplectic form, these are seen in the boundary integral defining the charge.

Finally, recall as mentioned above that \eqref{charge as boundary integral} does not guarantee that the finite charge associated to the infinitesimal expression $\delta\H[\Phi,\epsilon]$ does actually exist. For this, an additional integrability condition is necessary, and can be obtained from the requirement that one must have $\delta^2\H[\Phi,\epsilon]=0$ if $\slashed{\delta}\H[\Phi,\epsilon]=\delta\H[\Phi,\epsilon]$ is integrable.

\subsection{Boundary conditions and degrees of freedom}
\label{sec:2.3}

In the previous two subsection we have seen how, starting from the potential, one can define a conserved pre-symplectic form and derive an expression for (the variation of) the generators of infinitesimal gauge transformations and the associated charges. We would now like to explain how the potential is related to the fixation of boundary conditions and to the appearance of boundary degrees of freedom.

Using \eqref{variation of Lagrangian}, the variational principle for the action is
\be
\delta\int_ML[\Phi]=\int_ME[\Phi]\wedge\delta\Phi+\int_{\partial M}\theta[\Phi,\delta\Phi].
\ee
The spacetime boundary integral has to vanish in order for this variation to give the bulk equations of motion. For this, one can impose global boundary conditions which ensure the vanishing of the whole boundary integral. Alternatively, one can achieve this by imposing local boundary conditions, which amounts to finding a field configuration $\Phi\big|_{\partial M}=\Phi_\circ$ such that
\be\label{vanishing potential}
\de\theta_b[\Phi_\circ,\delta\Phi_\circ]=0.
\ee
Typically the choice of $\Phi_\circ$ is imposed by the physics or the solution under consideration, so instead of seeing \eqref{vanishing potential} as an equation for $\Phi_\circ$ one has to see it as an equation for the boundary term $b[\Phi]$. More precisely, knowing what field configuration $\Phi_\circ$ is to be fixed on the boundary, one looks for a boundary term which is then such that \eqref{vanishing potential} is satisfied. We will shortly encounter example illustrating this familiar matter.

Now, consider the action $\E^*\Phi$ of finite and \textit{field-dependent} gauge transformations on the fields $\Phi$, which possibly changes the Lagrangian by a total derivative as\footnote{Notice that this transformation rule is not satisfied in the case of non-Abelian Chern--Simons theory, where an additional bulk piece appears under finite gauge transformations. We comment on this in appendix \ref{appendix:CS}.}
\be
L[\E^*\Phi]=L[\Phi]+\de m[\Phi,\E],
\ee
and also possibly affects its boundary term as
\be\label{gauge transformation of boundary term}
b[\E^*\Phi]=b[\Phi]+\tilde{b}[\Phi,\E].
\ee
This is therefore of the general form
\be\label{gauge transformation of L}
L_b[\E^*\Phi]=L_b[\Phi]+\de m_b[\Phi,\E],
\ee
where
\be
m_b[\Phi,\E]\coloneqq m[\Phi,\E]+\tilde{b}[\Phi,\E]
\ee
contains a contribution from the non-invariance of the ``bare'' Lagrangian and a contribution from the non-invariance of the boundary term. For the sake of generality, we allow for $m_b[\Phi,\E]$ to be non-vanishing on-shell. Taking the variation of \eqref{gauge transformation of L} leads to
\be\label{gauge transformation spoiling variational principle}
\delta(L_b[\E^*\Phi])=E[\Phi]\wedge\delta\Phi+\de(\theta_b[\Phi,\delta\Phi]+\delta m_b[\Phi,\E])\stackrel{\text{bc}}{=}E[\Phi]\wedge\delta\Phi+\de\delta m_b[\Phi_\circ,\E],
\ee
where for the second equality we have used the boundary conditions \eqref{vanishing potential}. One can see that the field configuration $\Phi_\circ$ on the boundary leaves a contribution of the form $\de\delta m_b[\Phi_\circ,\E]$. As we shall see on concrete examples below, this contribution will always be non-vanishing in the case of field-dependent gauge transformations, thereby spoiling the well-posedness of the variational principle.

Now, it can happen that this boundary contribution vanishes for field-independent gauge transformations. In this case, the requirement of field-independence is enough to regain a well-defined variational principle. However, it can also be that even for field-independent transformations the boundary contribution is not vanishing. In this case, further restrictions have to be imposed at the boundary. The way to do so without further constraining the boundary fields $\Phi\big|_{\partial M}$ is to restrict the gauge transformations on the boundary to be such that
\be
\de\delta m_b[\Phi_\circ,\E_\circ]=0.
\ee
This can either be done by choosing gauge transformations which vanish on the boundary, or more generally by imposing a boundary equation on $\E$.

Now, recall that gauge transformations play the role of identifying field configurations which would have otherwise been thought of as physically distinct. Therefore, if we are led to restricting the allowed gauge transformations on the boundary in order to ensure the definiteness of the variational principle, it means that some of these configurations do not become identified anymore, or in other words that gauge degrees of freedom have now become physical degrees of freedom. This explains formally the origin of the ``would-be-gauge'' boundary degrees of freedom. We would of course like to know how to describe the dynamics of these boundary excitations. This can be achieved by constructing a gauge-invariant Lagrangian.

In order to make the Lagrangian $L_b[\Phi]$ gauge-invariant, one can add new fields $\Psi$ through a boundary term $\ell[\Phi,\Psi]$ transforming as
\be\label{gauge transformation of boundary term with new fields}
\ell[\E^*\Phi,\E^*\Psi]=\ell[\Phi,\Psi]+\tilde{\ell}[\Phi,\Psi,\E].
\ee
Indeed, if we find a boundary term such that
\be
\tilde{\ell}[\Phi,\Psi,\E]=-m_b[\Phi,\E]+\de(\text{something}),
\ee
then the Lagrangian
\be\label{general extended Lagrangian}
L_{b,\ell}[\Phi,\Psi]\coloneqq L_b[\Phi]+\de\ell[\Phi,\Psi]
\ee
is strictly gauge-invariant. Then we have
\be
\delta L_{b,\ell}[\Phi,\Psi]=\delta(L_{b,\ell}[\E^*\Phi,\E^*\Psi])=E[\Phi]\wedge\delta\Phi+\de(\theta_b[\Phi,\delta\Phi]+\delta\ell[\Phi,\Psi]),
\ee
so the variational principle is well-defined even when performing gauge-transformations. More precisely, it is well-defined when \eqref{vanishing potential} is satisfied and when the boundary fields $\Psi$ obey their boundary equations of motion.

In the literature, one can find two different explanations for the origin of the boundary degrees of freedom in e.g. Abelian Chern--Simons theory. Indeed, depending on the choice of boundary conditions and boundary term, it is either argued that boundary degrees of freedom arise because of the need to further constrain the field-independent gauge transformations on the boundary \cite{Tong:2016kpv}, or that they arise because the boundary Lagrangian which is added to obtain gauge-invariance has a kinetic term (which also means that $\de\delta m_b[\Phi_\circ,\E]$ contains $(\partial\E)^2$) \cite{Carlip:1994gy}. This state of affairs is not fully satisfactory, since it is not clear whether the boundary degrees of freedom arise because of the restriction on the allowed gauge transformations or because of the requirement of gauge-invariance. Furthermore, it is known that theories which are strictly gauge-invariant, i.e. with $m[\Phi,\E]=0$, may still have boundary degrees of freedom, although in this case none of the above arguments apply. Finally, notice that when adding boundary degrees of freedom through a boundary term $\ell[\Phi,\Psi]$, this latter will drop from the pre-symplectic current according to what we have discussed above. It is therefore clear that the covariant Hamiltonian method cannot tell us information about the boundary degrees of freedom introduced in this way.

Notice that here we reach this conclusion if indeed we think of the new degrees of freedom $\Psi$ as being added via a total divergence term $\ell[\Phi,\Psi]$ to the initial Lagrangian so as to make it gauge-invariant. One could however think of adding degrees of freedom $\Psi$ which are purely supported on the corner, and not on all of $\partial M$. This could be done by viewing $\ell[\Phi,\Psi]$ as an independent Lagrangian, computing its variation to extract its pre-symplectic potential, and then adding this potential as a corner term to the potential $\theta[\Phi,\delta\Phi]$. This is what has been done for example in \cite{Freidel:2016bxd}. However, it is rather immediate to see (and we will illustrate this with Abelian Chern--Simons theory) that the corner term which is thereby added to the potential is not enough to make this latter gauge-invariant and therefore pursue the construction of the extended gauge-invariant phase space along the lines of \cite{Donnelly:2016auv}. Again, this is a manifestation of our claim that considering the gauge-invariant Lagrangian is not enough in order to construct the extended phase space which contains the dressing boundary degrees of freedom leading to the dressed observables.

We would therefore like to find a more systematic criterion for investigating the presence and the nature of possible boundary degrees of freedom. Following the insight of \cite{Donnelly:2016auv}, this can be done by inspecting closer the properties of the pre-symplectic potential itself.

\subsection{Finite field-dependent transformations of the pre-symplectic potential}

In this subsection, we would like to know how the potential transforms under finite and field-dependent gauge transformations. For this, let us forget momentarily about possible boundary terms $\ell[\Phi,\Psi]$. Let us start by computing the gauge transformation of the variation of the Lagrangian. This is given by
\be\label{transformation of variation of Lagrangian}
(\delta L_b)[\E^*\Phi]
&=E[\E^*\Phi]\wedge\delta(\E^*\Phi)+\de\theta_b[\E^*\Phi,\delta(\E^*\Phi)]\nn\\
&=E[\Phi]\wedge\delta\Phi+\de p[\Phi,\E,\delta\E]+\de\theta_b[\E^*\Phi,\delta(\E^*\Phi)],
\ee
which is a generalization of Noether's result \eqref{setup for Noether's second} to the case of finite and field-dependent gauge transformations. We will see below on concrete examples that this relation is indeed correct. Notice, as suggested by the notation, that this relation involves first computing the variation $\delta$ of the Lagrangian, and then evaluating this variational quantity on the gauge-transformed fields $\E^*\Phi$. Now, demanding that $(\delta L_b)[\E^*\Phi]=\delta(L_b[\E^*\Phi])$ shows that we must necessarily have
\be\label{transformation of potential}
\theta_b[\E^*\Phi,\delta(\E^*\Phi)]=\theta_b[\Phi,\delta\Phi]+\de c[\Phi,\E,\delta]+\delta m_b[\Phi,\E]-p[\Phi,\E,\delta\E],
\ee
where the corner contribution can take the form
\be
\de c[\Phi,\E,\delta]=\de c_1[\Phi,\E,\delta\Phi]+\de c_2[\Phi,\E,\delta\E].
\ee
This simple result shows that the finite field-dependent gauge transformations of the potential can always be written in the form \eqref{transformation of potential}, and that by doing so one can identify the corner contribution. This generalizes of course straightforwardly to accommodate for the presence of a boundary term $\ell[\Phi,\Psi]$. More importantly, this shows that the potential is not gauge-invariant (even if $\delta\E=0$).

The only freedom in \eqref{transformation of potential} which could be used to cancel the corner ambiguity is that of playing with the boundary term $b[\Phi]$, adding boundary terms and fields $\ell[\Phi,\Psi]$, and thereby changing $\delta m_{b,\ell}[\Phi,\Psi,\E]$. More precisely, if the boundary terms of the Lagrangian transform as \eqref{gauge transformation of boundary term} and \eqref{gauge transformation of boundary term with new fields}, we have
\be
\delta m_{b,\ell}[\Phi,\Psi,\E]=\delta m[\Phi,\E]+\delta\tilde{b}[\Phi,\E]+\delta\tilde{\ell}[\Phi,\Psi,\E],
\ee
where $\delta m[\Phi,\E]$ is the contribution which comes from the bulk Lagrangian and cannot be adjusted. By changing the boundary terms, we can change the last two terms, but it is clear that there can be corner contributions in \eqref{transformation of potential} which cannot be absorbed by these last two terms, no matter what the boundary terms are chosen to be.

In these expressions one can see that $p[\Phi,\E,\delta\E]$ must necessarily contain the equations of motion (and therefore vanish on-shell) in order for the pre-symplectic current
\be
\omega[\delta_1,\delta_2,\Phi,\E]\coloneqq\delta\theta_b[\E^*\Phi,\delta(\E^*\Phi)]
\ee
to be closed on-shell. In fact, if $p[\Phi,\E,\delta\E]=p[E,\Phi,\E,\delta\E]\simeq0$ then we have that
\be\label{closedness and p term}
\de\omega[\delta_1,\delta_2,\Phi,\E]\simeq\de\omega[\Phi,\delta_1,\delta_2]\simeq0.
\ee
We will see that this condition is satisfied for all the examples treated in this paper, namely (Abelian and non-Abelian) Chern--Simons theory as well as three-dimensional gravity without a cosmological constant.

In non-Abelian Chern--Simons theory however, we will see in appendix \ref{appendix:CS} that there is an additional term in the transformed potential which is not vanishing on-shell and which comes from the behavior of the Lagrangian under finite gauge transformations (see previous footnote). It is only when computing the variation of this term, i.e. when going to the pre-symplectic current, that it will become a total exterior derivative.

\subsection{Extended pre-symplectic potential}

Let us now consider the general case where we have boundary terms $b[\Phi]$ and $\ell[\Phi,\Psi]$. This general Lagrangian has a potential given by $\theta_{b,\ell}[\Phi,\Psi,\delta\Phi,\delta\Psi]$ and which under a gauge transformation becomes
\be\label{general theta b l gauge transformed}
\theta_{b,\ell}[\E^*\Phi,\E^*\Psi,\delta(\E^*\Phi),\delta(\E^*\Psi)]=\theta_{b,\ell}[\Phi,\Psi,\delta\Phi,\delta\Psi]+\de c[\Phi,\E,\delta]+\delta m_{b,\ell}[\Phi,\Psi,\E]-p[E,\Phi,\E,\delta\E].
\ee
The gauge-non-invariance of the potential can be corrected by introducing new fields $\Pi$. Generally, one can obtain off-shell gauge-invariance by considering
\be\label{fully extended potential}
\theta_{b,\ell,c,m,p}[\Phi,\Psi,\Pi,\delta\Phi,\delta\Psi,\delta\Pi]\coloneqq\theta_{b,\ell}[\Phi,\Psi,\delta\Phi,\delta\Psi]+\de c[\Phi,\Pi,\delta]+\delta m_{b,\ell}[\Phi,\Psi,\Pi]-p[E,\Phi,\Pi,\delta\Pi]
\ee
provided we choose carefully how the fields $\Pi$ transform under the action of $\E^*$.

It should already be clear from the above formula that considering the extended potential and the extended Lagrangian \eqref{general extended Lagrangian} is not at all equivalent. Indeed, only the former knows about the corner ambiguities. This will be seen on concrete examples in the next two sections.

Notice that the expression \eqref{fully extended potential} is in a sense a ``minimal'' gauge-invariant extension of the potential, which is the same as the right-hand side of \eqref{general theta b l gauge transformed} where we have replaced $\E$ by $\Pi$. We shall see explicitly on the examples considered in the rest of this work how a gauge-invariant expression like \eqref{fully extended potential} can indeed be obtained by computing \eqref{general theta b l gauge transformed} and then promoting the parameters $\E$ to new fields $\Pi$. The expression \eqref{fully extended potential} is however not unique, and one is free to add any functional of the fields and their variations as long as this functional itself is invariant under the finite transformations $\E^*$. One can also consider adding contributions from new fields other than $\Pi$. We will not consider the effect of these additional ambiguities in this work. However, let us simply mention that when allowing for such extra ambiguities, one should do so in a manner which preserves the properties of gauge-invariance which are gained from considering the extended potential. This is of course the aforementioned gauge-invariance property under $\E^*$, and, as we shall see later on on concrete examples, the fact that generator of gauge transformations computed from the extended symplectic structure are vanishing on-shell. We expect that once these conditions related to gauge-invariance are satisfied, the allowed terms that one can add to \eqref{fully extended potential} could be related to the dynamics of the boundary degrees of freedom (i.e. describe the canonically-conjugated momenta to the fields $\Pi$). We postpone the investigation of such a possibility to future work.

By construction, the extended gauge-invariant potential is invariant under infinitesimal gauge transformations, i.e.
\be\label{infinitesimal transformation of extended theta}
\theta_{b,\ell,c,m,p}[\Phi,\Psi,\Pi,\delta_\epsilon\Phi,\delta_\epsilon\Psi,\delta_\epsilon\Pi]=0.
\ee
Furthermore, we will see that the term involving the equations of motion is such that
\be
p[E,\Phi,\Pi,\delta_\epsilon\Pi]=P[E,\Phi,\epsilon],
\ee
where $P[E,\Phi,\epsilon]$ is the quantity appearing \eqref{setup for Noether's second} and \eqref{Noether in terms of Q and P}. Notice that in this equality the new fields $\Pi$ have actually dropped from the right-hand side. This is due to the way in which the fields $\Pi$ have been chosen to transform under the infinitesimal gauge transformations $\delta_\epsilon$. We shall encounter below multiples examples illustrating this property. Now, in order to understand the consequence of relation \eqref{infinitesimal transformation of extended theta}, let us focus on a Lagrangian which has no boundary terms and is strictly gauge-invariant, i.e. with $b[\Phi]=\ell[\Phi,\Psi]=m[\Phi,\epsilon]=0$. In this case \eqref{infinitesimal transformation of extended theta} leads to
\be
\theta[\Phi,\delta_\epsilon\Phi]+\de c[\Phi,\delta_\epsilon]=P[E,\Phi,\epsilon].
\ee
Using this corner ambiguity in the definition \eqref{Noether current} of the Noether current and comparing with \eqref{Noether in terms of Q and P} shows that the associated Noether charge is actually vanishing. This is a surprising result, which indicates that there is a systematic way of choosing the corner ambiguity appearing in the definition of the Noether current in such a way as to obtain a vanishing Noether charge for gauge symmetries. 

Now, from this extended potential we can construct a pre-symplectic current. According to \eqref{pre-symplectic current}, this latter will not depend on the boundary terms $b[\Phi]$ and $\ell[\Phi,\Psi]$, but will contain the corner terms. More explicitly, we have
\be
\omega_{b,\ell,c,m,p}[\Phi,\Psi,\Pi,\delta_1,\delta_2]
&=\omega[\Phi,\delta_1,\delta_2]+\de w[\Phi,\Pi,\delta_1,\delta_2]-\delta p[E,\Phi,\Pi,\delta\Pi]=\omega_{c,p}[\Phi,\Pi,\delta_1,\delta_2],
\ee
where we have denoted the corner contribution by
\be
\de w[\Phi,\Pi,\delta_1,\delta_2]\coloneqq\de\delta c[\Phi,\Pi,\delta].
\ee
One can see that the pre-symplectic current does not depend on the extra boundary fields $\Psi$. Now, since we focus on the case of theories for which $p[E,\Phi,\Pi,\delta\Pi]\simeq0$, this pre-symplectic current is closed on-shell, i.e.
\be
\de\omega_{c,p}[\Phi,\Pi,\delta_1,\delta_2]\simeq\de\omega_c[\Phi,\Pi,\delta_1,\delta_2]=\de\omega[\Phi,\delta_1,\delta_2]\simeq0.
\ee
In \eqref{integration of omega} we have already discussed the integration of $\de\omega[\Phi,\delta_1,\delta_2]$. Instead, integrating the $(d,2)$-form $\de\omega_c[\Phi,\Pi,\delta_1,\delta_2]$ over $N$ leads to
\be
\int_N\de\omega_c[\Phi,\Pi,\delta_1,\delta_2]=\int_{\partial N}\omega_c[\Phi,\Pi,\delta_1,\delta_2]=\left(\int_{\Sigma_2}-\int_{\Sigma_1}+\int_B\right)\omega_c[\Phi,\Pi,\delta_1,\delta_2]\simeq0.
\ee
Assuming that the symplectic flux coming from $B$ is vanishing, we get that the pre-symplectic $(0,2)$-form defined as
\be\label{conserved form with corner fields}
\Omega_{\Sigma,\partial\Sigma}[\Phi,\Pi,\delta_1,\delta_2]
&=\int_\Sigma\omega_c[\Phi,\Pi,\delta_1,\delta_2]\nn\\
&=\int_\Sigma\omega[\Phi,\delta_1,\delta_2]+\int_{\partial\Sigma}w[\Phi,\Pi,\delta_1,\delta_2]\nn\\
&=\Omega_\Sigma[\Phi,\delta_1,\delta_2]+\Omega_{\partial\Sigma}[\Phi,\Pi,\delta_1,\delta_2]
\ee
does not depend on the hypersurface $\Sigma$ when the equations of motion are satisfied. Now, notice that the condition of vanishing flux at $B$ translates into
\be
\int_B\omega[\Phi,\delta_1,\delta_2]=-\int_B\de w[\Phi,\Pi,\delta_1,\delta_2],
\ee
and is therefore a generalization of the condition which leads to \eqref{conserved form with corner}. The fields $\Pi$ which we have introduced allow to compensate for the ``leaking'' of the pre-symplectic form through $B$. Moreover, they give a boundary contribution the pre-symplectic form.

We are now going to study the consequences of this construction through the examples of Abelian Chern--Simons theory and first order three-dimensional gravity.

\section{Abelian Chern--Simons theory}
\label{sec:3}

In this section, we study the boundary degrees of freedom of Abelian Chern--Simons theory in three ways. First, we are going to review the standard Lagrangian viewpoint. This will illustrate the ambiguities which we have discussed on general grounds in section \ref{sec:2.3}. Then, we will discuss the Hamiltonian treatment along the lines of Regge and Teitelboim. This will enable us to identify the boundary observables and their algebra, but we will see that this requires to impose that the parameters of gauge transformations have compact support. Finally, we will see following the construction of \cite{Donnelly:2016auv} that the introduction of new boundary fields allows to relax this condition and to disentangle the role of generators of gauge transformations from that of boundary observables. More precisely, we will obtain boundary observables and at the same time have a generator of gauge transformations which is vanishing on-shell (and as such has no Hamiltonian charge).

\subsection{Lagrangian}

Let us consider the Lagrangian
\be\label{gauge-non-invariant Abelian CS}
L[A]=A\wedge\de A.
\ee
Its variation is
\be
\delta L[A]=2\delta A\wedge\de A+\de\theta[A,\delta A]=2\delta A\wedge\de A+\de(\delta A\wedge A).
\ee
Under infinitesimal and finite gauge transformations, the fields transform as
\be
\delta_\alpha A=\de\alpha,\q\delta_\alpha F=0,\q\alpha^*A=A+\de\alpha,\q\alpha^*F=F,
\ee
where $F=\de A$ is the Abelian field strength. The Lagrangian, on the other hand, transforms as
\be\label{infinitesimal transformation of Abelian CS}
\delta_\alpha L[A]=\de(A\wedge\de\alpha)=\de(\alpha\de A)\simeq0,
\ee
and
\be\label{finite transformation of Abelian CS}
L[\alpha^*A]=L[A]+\de(A\wedge\de\alpha)=L[A]+\de(\alpha\de A)\simeq L[A].
\ee
We are now going to discuss how this transformation of the Lagrangian can interfere with the variational principle and the boundary conditions, and how this is in turn related to the appearance of a boundary dynamics.

Note that the specific issue of deriving the boundary dynamics is beyond the scope of the present work, which is simply interested in the construction of the extended gauge-invariant phase space for Chern--Simons theory and three-dimensional gravity. However, since Abelian Chern--Simons theory is the simplest and quintessential example of a theory for which the boundary dynamics can easily be obtained from Lagrangian considerations, we find it interesting to reproduce carefully the arguments for the sake of completeness. In fact, we will see that this discussion requires to go through additional subtleties related to the choice of boundary conditions. In the following two examples, we therefore discuss the boundary dynamics of \eqref{gauge-non-invariant Abelian CS} for two choices of boundary conditions. The reader who is interested in the Hamiltonian description of the boundary observables can safely skip to section \ref{subsec:Hamiltonian Abelian CS}.

\subsubsection{Example 1}

With coordinates $x^\mu=(x^0,x^1,x^2)=(t,\phi,r)$, we have the on-shell variation\footnote{Here and in the rest of the text we use a slight abuse of notation, and write the on-shell equality $\simeq$ even though we have not yet imposed the boundary conditions that cancel the boundary contribution to the variation and thereby make the bulk equations of motion well-defined. This is just a simple way of dropping bulk contributions proportional to the yet-to-be-defined equations of motion.}
\be\label{boundary variation Abelian CS}
\delta S[A]\simeq\int_{\partial M}(\delta A_tA_\phi-\delta A_\phi A_t).
\ee
One way to cancel this boundary variation without the need to introduce a boundary term is to set $(A_t-vA_\phi)\big|_{\partial M}=0$, where we have allowed for the presence of a free parameter $v\in\mathbb{R}$ (in the fractional Hall effect, this corresponds to the velocity of the bosonic excitations on the boundary \cite{Wen:1992vi,Tong:2016kpv}). Using this boundary condition and the gauge transformation \eqref{finite transformation of Abelian CS}, we get
\be
\delta(S[\alpha^*A])\simeq\delta\int_{\partial M}A_\phi(v\partial_\phi\alpha-\partial_t\alpha),
\ee
which is not vanishing even in the case $\delta\alpha=0$ of field-independent gauge transformations. This is an illustration of the general equation \eqref{gauge transformation spoiling variational principle} which we have discussed in the previous section. One way to cancel this new boundary variation without further restricting the boundary connection (since $\delta A_\phi\big|_{\partial M}=0$ would imply $\delta A_t\big|_{\partial M}=0$ and freeze the dynamics) is to choose one of the two following restrictions on the gauge fields:
\be
\alpha\big|_{\partial M}=0,\q(\partial_t\alpha-v\partial_\phi\alpha)\big|_{\partial M}=0.
\ee
As discussed in \ref{sec:2.3}, this gives rise to boundary would-be-gauge degrees of freedom. However, the exact dynamics of these degrees of freedom is still unclear at the moment. This suggests that we look at the gauge-invariant action.

The gauge-invariant action is obtained simply by promoting $\alpha$ to a dynamical field $a$ transforming as $\alpha^*a=a-\alpha$, which gives
\be\label{gauge-invariant action of example 1}
S_\ell[A,a]\coloneqq S[A]+\int_{\partial M}(A_t\partial_\phi a-A_\phi\partial_ta).
\ee
However, this action has no kinetic term for $a$. One could be tempted to derive equations of motion by computing the functional variation with respect to $A_\phi\big|_{\partial M}$, but this is not well-defined since $\delta A_\phi$ appears in the bulk. One procedure sometimes followed in the literature is to write the gauge field as a gauge transformation $A_\mu=\tilde{A}_\mu+\partial_\mu\alpha$ \cite{Dunne:1998qy,Tong:2016kpv,Wen:1992vi}. Using the boundary condition to write $\tilde{A}_t\big|_{\partial M}=v(\tilde{A}_\phi+\partial_\phi\alpha)-\partial_t\alpha$, the action then becomes
\be
S[\tilde{A},\alpha]=S[\tilde{A}]+\int_{\partial M}(v\partial_\phi\alpha-\partial_t\alpha)(\tilde{A}_\phi+\partial_\phi\alpha).
\ee
Choosing the gauge $\tilde{A}_\phi=\tilde{A}_r=0$ in the bulk (which solves the constraint $\partial_\phi A_r-\partial_rA_\phi=0$ conjugated to $A_t$) then leads to the so-called Floreanini--Jackiw action
\be
S[\alpha]\coloneqq\int_{\partial M}(v\partial_\phi\alpha-\partial_t\alpha)\partial_\phi\alpha.
\ee
The equations of motion describing the boundary dynamics of the would-be-gauge degrees of freedom are
\be\label{equation of motion for a}
\partial_t\partial_\phi\alpha-v\partial^2_\phi\alpha=0,
\ee
or, introducing the field $\rho\coloneqq\partial_\phi a$,
\be\label{equation of motion for rho}
\partial_t\rho-v\partial_\phi\rho=0.
\ee
The solution is given by a chiral wave $\rho(\phi+vt)$.

We have seen in this example that it is possible to obtain a boundary Lagrangian describing the dynamics of $\alpha$. Now, let us slightly modify the boundary conditions in order to see how the equations of motion \eqref{equation of motion for a} or \eqref{equation of motion for rho} can be regained.

\subsubsection{Example 2}

If on the boundary we wish to fix $\delta(A_t-vA_\phi)\big|_{\partial M}=0$, we have to add a boundary term to the action and consider
\be\label{chiral boundary term 1}
S_{b_1}[A]\coloneqq S[A]+\int_{\partial M}A_\phi(A_t-vA_\phi),
\ee
which is indeed such that
\be
\delta S_{b_1}[A]\simeq2\int_{\partial M}A_\phi\delta(A_t-vA_\phi).
\ee
Under a finite gauge transformation we get
\be
S_{b_1}[\alpha^*A]=S_{b_1}[A]+\int_{\partial M}\partial_\phi\alpha\big(\partial_t\alpha-v\partial_\phi\alpha+2(A_t-vA_\phi)\big),
\ee
which implies that
\be
\delta(S_{b_1}[\alpha^*A])\simeq2\int_{\partial M}A_\phi\delta(A_t-vA_\phi)+\delta\int_{\partial M}\partial_\phi\alpha\big(\partial_t\alpha-v\partial_\phi\alpha+2(A_t-vA_\phi)\big).
\ee
This has to be vanishing in order for the variational principle to be well-defined. The first term is vanishing with our initial choice of boundary conditions. The second term will be vanishing if $\delta\alpha=0$. If we insist on considering field-dependent gauge parameters, then these have to be restricted to satisfy one of the two following two conditions:
\be
\partial_\phi\alpha\big|_{\partial M}=0,\q\big(\partial_t\alpha-v\partial_\phi\alpha+2(A_t-vA_\phi)\big)\big|_{\partial M}=0.
\ee
Again, it is now still not clear what the dynamics of the boundary degrees of freedom is.

However, looking at the gauge-invariant action obtained by promoting $\alpha$ to be a dynamical field $a$, one gets
\be\label{Abelian WZNW action}
S_{b_1,\ell_1}[A,a]\coloneqq S_{b_1}[A]+\int_{\partial M}\partial_\phi a\big(\partial_ta-v\partial_\phi a+2(A_t-vA_\phi)\big),
\ee
which contains a kinetic term for $a$ (at the difference with \eqref{gauge-invariant action of example 1}). This is actually nothing but the Abelian WZNW action for the fields $a$ coupled to $A$ \cite{Carlip:1994gy} (although this latter is usually written in terms of complex coordinates on $\partial M$). The corresponding equations of motion can be derived from
\be
\delta S_{b_1,\ell_1}[A,a]
&\simeq2\int_{\partial M}A_\phi\delta(A_t-vA_\phi)+\delta\int_{\partial M}\partial_\phi a\big(\partial_ta-v\partial_\phi a+2(A_t-vA_\phi)\big)\nn\\
&=\int_{\partial M}\Big(\partial_\phi\delta a\big(\partial_ta-v\partial_\phi a+2(A_t-vA_\phi)\big)+\partial_\phi a(\partial_t\delta a-v\partial_\phi\delta a)\Big),
\ee
where we have used the boundary condition $\delta(A_t-vA_\phi)\big|_{\partial M}=0$. The equation of motion for $a$ is therefore
\be
\partial_t\partial_\phi a-v\partial^2_\phi a=0,
\ee
which is the same as \eqref{equation of motion for a}. This therefore shows that we have obtained the same boundary dynamics as in the previous example, but the manipulations of the action which are involved in this derivation are completely different, and this can be traced back to the difference in the choice of boundary conditions.

Now, notice that instead of the boundary term in \eqref{chiral boundary term 1} one could have also chosen
\be
S_{b_2}[A]\coloneqq S[A]+\f{1}{2v}\int_{\partial M}(A_t+vA_\phi)(A_t-vA_\phi),
\ee
which is such that
\be
\delta S_{b_2}[A]\simeq\int_{\partial M}A_\phi\delta(A_t-vA_\phi),
\ee
and therefore enables us to fix the same boundary conditions. Now, following the same steps as above one finds that the gauge-invariant action is given by
\be
S_{b_2,\ell_2}[A,a]\coloneqq S_{b_2}[A]+\f{1}{2v}\int_{\partial M}(\partial_ta+v\partial_\phi a)\big(\partial_ta-v\partial_\phi a+2(A_t-vA_\phi)\big),
\ee
and the equation of motion for $a$ is therefore
\be
\partial^2_ta-v^2\partial^2_\phi a=0,
\ee
or, introducing the field $\rho\coloneqq\partial_ta+v\partial_\phi a$,
\be
\partial_t\rho-v\partial_\phi\rho=0.
\ee
The above two simple examples illustrate how the Lagrangian derivation of the boundary dynamics of Abelian Chern--Simons theory does actually depend on how the boundary conditions are written. Although being a simple fact, this also illustrate how, depending on the choice of boundary conditions, field-dependent or field-independent gauge-transformations can interfere with the variational principle. Most importantly, it shows that there is no uniquely defined procedure which enables for the derivation of the boundary dynamics.

Let us now turn to the Hamiltonian description of the boundary observables. As is well-known, these satisfy a current algebra which corresponds to the symplectic structure of the Abelian WZNW action \eqref{Abelian WZNW action} \cite{Gawedzki:2001ye,Gawedzki:2001rm,Papadopoulos:1992rs}.

\subsection{Hamiltonian}
\label{subsec:Hamiltonian Abelian CS}

We now review how boundary observables arise in relation with the requirement of functional differentiability of the constraints of the Hamiltonian framework \cite{Regge:1974zd,Balachandran:1995qa}. The Hamiltonian action is
\be\label{Hamiltonian Abelian CS action}
S[A]=\int_\mathbb{R}\de t\int_\Sigma\big(\partial_0A\wedge A+2A_0\de A-\de(AA_0)\big),
\ee
where the differential forms are understood as pulled-back to the spatial slice $\Sigma$. The canonical Poisson bracket is $\lb A_a(x),A_b(y)\rb=-\teps_{ab}\delta^2(x,y)/2$, and generic brackets are given by
\be
\lb f_1,f_2\rb=-\f{1}{2}\teps_{ab}\int_\Sigma\de^2x\int_\Sigma\de^2y\,\delta^2(x,y)\f{\delta f_1}{\delta A_a(x)}\f{\delta f_2}{\delta A_b(y)}.
\ee
Let us now consider the smeared flatness constraint\footnote{The action \eqref{Hamiltonian Abelian CS action} is of the form $p\dot{q}-\mathcal{H}_\text{tot}$, where $\mathcal{H}_\text{tot}$ is the total Hamiltonian which contains the primary constraints. Although we could discard the factor $-2$ for simplicity, we choose to keep it in order to ensure that all the quantities computed throughout this section are in exact agreement.}
\be\label{Abelian flatness constraint}
\F[\alpha]\coloneqq-2\int_\Sigma\alpha\de A\simeq0,
\ee
Its variation is given by
\be\label{variation of abelian CS flatness constraint}
\delta\F[\alpha]=-2\int_\Sigma(\delta\alpha\de A+\delta A\wedge\de\alpha)-2\int_{\partial\Sigma}\alpha\delta A.
\ee
In order to compute the Poisson bracket between the constraint and any other function on phase space, these should be functionally differentiable. In the case of the flatness constraint, this requires one of the following conditions:
\begin{itemize}
\item[(1)]$\q$defining an extended generator by $\displaystyle\slashed{\delta}\F_c[\alpha]\coloneqq\delta\F[\alpha]+2\int_{\partial\Sigma}\alpha\delta A$,
\item[(2)]$\q$considering parameters $\bar{\alpha}$ with compact support, i.e. such that $\bar{\alpha}\big|_{\partial\Sigma}=0$,
\item[(3)]$\q$imposing $\delta A\big|_{\partial\Sigma}=0$,
\item[(4)]$\q$imposing $\alpha\delta A\big|_{\partial\Sigma}=0$,
\item[(5)]$\q$imposing the global condition $\displaystyle\int_{\partial\Sigma}\alpha\delta A=0$.
\end{itemize}
Let us start with the least restrictive choice and walk our way towards the derivation of the boundary observables.

Condition (1) is the least restrictive choice since it does not require imposing any conditions on the fields or the gauge parameter at the boundary $\partial\Sigma$. In particular, if we allow for field-dependent parameters $\alpha$, i.e. with $\delta\alpha\neq0$, the variation inside of the boundary contribution cannot be pulled outside of the integral. An important case of field-dependent transformations is for spatial diffeomorphisms. These are obtained with $\alpha=\xi\ip A$. Indeed, in this case we get
\be
\lb\F_c[\xi\ip A],A\rb=\de(\xi\ip A)+\xi\ip\de A=\L_\xi A.
\ee
Assuming that $\delta\alpha=0$, we can compute the brackets
\be\label{Abelian F_c F_c algebra}
\lb\F_c[\alpha],A\rb=\de\alpha,\q\lb\F_c[\alpha],\F_c[\beta]\rb=2\int_{\partial\Sigma}\de\alpha\beta,
\ee
showing that $\F_c[\alpha]$ is indeed the generator of gauge transformations, and that its algebra is anomalous. Furthermore, the extended generator is then integrable and one can write
\be
\F_c[\alpha]=\F[\alpha]+2\int_{\partial\Sigma}\alpha A\simeq2\int_{\partial\Sigma}\alpha A.
\ee
Now, if we want the algebra \eqref{Abelian F_c F_c algebra} to close, we have to impose that the gauge parameters be vanishing at the boundary. In this case we are naturally led to condition (2), and we get back the original constraint $\F_c[\bar{\alpha}]=\F[\bar{\alpha}]\simeq0$ with a closed algebra
\be
\lb\F[\bar{\alpha}],\F[\bar{\beta}]\rb=0.
\ee

For an arbitrary smearing parameter $\alpha$ which does not vanish on $\partial\Sigma$, let us now consider the quantity
\be\label{Abelian CS corner observable}
\O[\alpha]\coloneqq-2\int_\Sigma A\wedge\de\alpha=\F[\alpha]+2\int_{\partial\Sigma}\alpha A\simeq2\int_{\partial\Sigma}\alpha A,
\ee
which is not vanishing (i.e. not a flatness constraint) since $\alpha$ does not have to satisfy (2). This is an observable since we have
\be
\lb\O[\alpha],\F[\bar{\alpha}]\rb=2\int_{\partial\Sigma}\de\alpha\bar{\alpha}=0,
\ee
by virtue of the fact that $\bar{\alpha}$ satisfies (2). Furthermore, for $\alpha$ and $\beta$ such that $\alpha\big|_{\partial\Sigma}=\beta\big|_{\partial\Sigma}$, we have $(\alpha-\beta)\big|_{\partial\Sigma}=0$,which implies that
\be
\O[\alpha]-\O[\beta]=\F[\alpha-\beta]\simeq0.
\ee
This shows that the observables $\O[\alpha]$ are located on $\partial\Sigma$. Finally, these observables satisfy the current algebra
\be
\lb\O[\alpha],\O[\beta]\rb=2\int_{\partial\Sigma}\de\alpha\beta.
\ee
We see that these observables arise not simply from the requirement of functional differentiability of the flatness constraint, but with the particular addition of condition (2) of compact support for the gauge parameters.

Let us summarize this section by recalling the roles played by $\F_c$, $\F$ and $\O$. We have first obtained a differentiable generator of the gauge transformations by defining the extended generator $\F_c$. This generator is not a constraint, and on the surface of the constraint \eqref{Abelian flatness constraint} it is equal to a surface term. Also, it does not form a closed algebra. We have then restricted ourselves to compactly-supported parameters $\bar{\alpha}$ of gauge transformations, for which $\F_c$ then agrees with the original contraint $\F$ and is differentiable. Then, because of the requirement of compact support for the parameters, we have seen that the quantity $\O$ is an observable since it has vanishing Poisson bracket with the constraint. As expected, for general parameters $\alpha$ which do not vanish on the boundary, one can see that $\F_c$ and $\O$ are actually weakly equal to the same surface integral and have the same Poisson bracket.

\subsection{Extended pre-symplectic potential}

In order to understand properly the construction of the gauge-invariant extended potential, we are going to study three Lagrangians for Abelian Chern--Simons theory. The first one is the gauge-non-invariant Lagrangian \eqref{gauge-non-invariant Abelian CS}, the second one is the gauge-invariant Lagrangian obtained by adding boundary fields following \eqref{general extended Lagrangian}, and the third one is a gauge-invariant Lagrangian obtained by adding bulk fields. This will illustrate a result which is already clear from the general analysis of the previous section, namely that having a gauge-invariant Lagrangian is not enough to guarantee the gauge-invariance of its potential. As anticipated and illustrated in \cite{Donnelly:2016auv}, this means in turn that the extended gauge-invariant potential encodes more information about the boundary degrees of freedom than the Lagrangian alone.

\subsubsection{Gauge-non-invariant Lagrangian}

The Lagrangian which we consider here is simply \eqref{gauge-non-invariant Abelian CS}, and we would like to show that we have the equality $\delta(L[\alpha^*A])=(\delta L)[\alpha^*A]$. This computation will involve looking at the gauge transformation of the potential. Computing the variation of \eqref{finite transformation of Abelian CS}, we obtain
\be
\delta(L[\alpha^*A])
&=2\delta A\wedge\de A+\de\big(\theta[A,\delta A]+\delta(A\wedge\de\alpha)\big)\nn\\
&=2\delta A\wedge\de A+\de\big(\theta[A,\delta A]+\delta(\alpha\de A)\big),
\ee
where the potential is given by
\be
\theta[A,\delta A]=\delta A\wedge A.
\ee
On the other hand, computing the gauge transformation of the variation of the Lagrangian leads to
\be
(\delta L)[\alpha^*A]=2\delta A\wedge\de A+2\de(\delta\alpha\de A)+\de\theta[\alpha^*A,\delta(\alpha^*A)],
\ee
where the gauge transformation of the potential is found to be
\begin{subequations}
\be
\theta[\alpha^*A,\delta(\alpha^*A)]
&=\theta[A,\delta A]+\de\big(\delta\alpha(2A+\de\alpha)\big)+\delta(A\wedge\de\alpha)-2\delta\alpha\de A\label{transformation of original Abelian CS potential a}\\
&=\theta[A,\delta A]+\de\big(\delta\alpha(A+\de\alpha)-\alpha\delta A\big)+\delta(\alpha\de A)-2\delta\alpha\de A.\label{transformation of original Abelian CS potential b}
\ee
\end{subequations}
The proof of this equation is given in appendix \ref{appendix:proofs}. This shows indeed that we have the equality $\delta(L[\alpha^*A])=(\delta L)[\alpha^*A]$, and that $\theta[\alpha^*A,\delta(\alpha^*A)]$ is of the general form given in \eqref{transformation of potential}.

Now, we can add extra fields in order to define an extended gauge-invariant potential. This can be done by considering a field $u$ which transforms as $\alpha^*u=u-\alpha$. Then, defining the extended potential
\begin{subequations}
\be
\theta_{c,m,p}[A,u,\delta A,\delta u]
&\coloneqq\theta[A,\delta A]+\de\big(\delta u(2A+\de u)\big)+\delta(A\wedge\de u)-2\delta u\de A\label{Abelian CS extended potential 1}\\
&\phantom{:}=\theta[A,\delta A]+\de\big(\delta u(A+\de u)-u\delta A\big)+\delta(u\de A)-2\delta u\de A,\label{Abelian CS extended potential 1bis}
\ee
\end{subequations}
a direct calculation shows that we indeed have the gauge-invariance property
\be
\theta_{c,m,p}[\alpha^*A,\alpha^*u,\delta(\alpha^*A),\delta(\alpha^*u)]=\theta_{c,m,p}[A,u,\delta A,\delta u].
\ee
We have therefore succeeded in defining an off-shell gauge-invariant potential for the Lagrangian \eqref{gauge-non-invariant Abelian CS}. This is one of the main results of the present article, namely the construction following \cite{Donnelly:2016auv} of the gauge-invariant potential for Abelian Chern--Simons theory.

\subsubsection{Gauge-invariant boundary-extended Lagrangian}
\label{subsubsec:Gauge-invariant boundary-extended Lagrangian}

As observed in \eqref{infinitesimal transformation of Abelian CS}, the Lagrangian \eqref{gauge-non-invariant Abelian CS} is not gauge-invariant. This can be remedied by introducing a new boundary field $a$ which transforms as $\alpha^*a=a-\alpha$. Then, one way of obtaining a gauge-invariant theory is to consider the Lagrangian
\be\label{boundary-extended Lagrangian 1}
L_{\ell_1}[A,a]\coloneqq L[A]+\de(A\wedge\de a).
\ee
Alternatively, one can consider
\be
L_{\ell_2}[A,a]\coloneqq L[A]+\de(a\de A).
\ee
Evidently, these two gauge-invariant Lagrangians lead to the same equations of motion, but differ by a corner ambiguity in the potential.

Let us now look at the the potentials for these Lagrangians. For the Lagrangian \eqref{boundary-extended Lagrangian 1} it is given by
\be
\theta_{\ell_1}[A,a,\delta A,\delta a]=\theta[A,\delta A]+\delta(A\wedge\de a).
\ee
Looking at gauge transformations, we get that
\be
\theta_{\ell_1}[\alpha^*A,\alpha^*a,\delta(\alpha^*A),\delta(\alpha^*a)]=\theta_{\ell_1}[A,a,\delta A,\delta a]+\de\big(\delta\alpha(2A+\de\alpha)\big)+\delta\de(\alpha\de a)-2\delta\alpha\de A,
\ee
which is of the form \eqref{transformation of potential}. When comparing this with \eqref{transformation of original Abelian CS potential a}, we can see that the term $\delta m[A,\alpha]=\delta(A\wedge\de\alpha)$ has been replaced by $\delta m_\ell[a,\alpha]=\delta\de(\alpha\de a)$. This is due to the fact that we have restored the gauge-invariance of the Lagrangian by adding a boundary term, but the gauge transformation of this boundary term leads to a corner term. More precisely, this means that we have added a boundary term $\ell[A,a]=A\wedge\de a$ transforming as
\be
\ell[\alpha^*A,\alpha^*a]
&=\ell[A,a]+\tilde{\ell}[A,\alpha]\nn\\
&=\ell[A,a]-m[A,\alpha]+\de(\text{something})\nn\\
&=A\wedge\de a-A\wedge\de\alpha+\de(\alpha\de a),
\ee
thereby leading in the potential to
\be
\delta m_\ell[a,\alpha]=\delta m[A,\alpha]+\delta\tilde{\ell}[A,\alpha]=\delta\de(\alpha\de a).
\ee
Let us now look at the second potential, i.e.
\be
\theta_{\ell_2}[A,a,\delta A,\delta a]=\theta[A,\delta A]+\delta(a\de A),
\ee
and compute its gauge-transformation. This is given by
\be
\theta_{\ell_2}[\alpha^*A,\alpha^*a,\delta(\alpha^*A),\delta(\alpha^*a)]
&=\theta_{\ell_2}[A,a,\delta A,\delta a]+\de\big(\delta\alpha(A+\de\alpha)-\alpha\delta A\big)-2\delta\alpha\de A.
\ee
When comparing this with \eqref{transformation of original Abelian CS potential b}, we can see that the term $\delta m[A,\alpha]=\delta(\alpha\de A)$ has now disappeared.

Now, the two potentials can also be made off-shell gauge-invariant, just like in the previous subsection. For this, we simply need to consider
\be\label{Abelian CS extended potential 2}
\theta_{\ell_1,c,m,p}[A,a,u,\delta A,\delta a,\delta u]\coloneqq\theta_{\ell_1}[A,a,\delta A,\delta a]+\de\big(\delta u(2A+\de u)\big)+\delta\de(u\de a)-2\delta u\de A,
\ee
or
\be\label{Abelian CS extended potential 3}
\theta_{\ell_2,c,p}[A,a,u,\delta A,\delta a,\delta u]\coloneqq\theta_{\ell_2}[A,a,\delta A,\delta a]+\de\big(\delta u(A+\de u)-u\delta A\big)-2\delta u\de A.
\ee
We can see here that the introduction of the boundary fields $a$, which have the role of making the Lagrangian gauge-invariant, is not enough in order to guarantee that the potential is gauge-invariant.

Furthermore, notice that one could also have considered the boundary Lagrangian $\ell[A,a]=a\de A$ (or equivalently $\ell[A,a]=A\wedge\de a$), computed its potential $a\delta A$, and added this potential as the corner ambiguity to the potential $\delta A\wedge A$ of the Chern--Simons Lagrangian. However, it is clear that this procedure does also not lead to the fully gauge-invariant potential which we have constructed.

\subsubsection{Gauge-invariant bulk-extended Lagrangian}

One other way of making the Lagrangian gauge-invariant is through the introduction of an additional bulk connection $B$ which transforms as $\alpha^*B=B+\de\alpha$. Then, one can consider
\be\label{bulk-extended Abelian CS}
L[A,B]\coloneqq(A-B)\wedge\de A,
\ee
whose variation is given by
\be
\delta L[A,B]=\delta A\wedge(2\de A-\de B)-\delta B\wedge\de A+\de\big(\delta A\wedge(A-B)\big),
\ee
showing that the combined equations of motion are\footnote{Note that the Lagrangian
$L'[A,B]\coloneqq(A-B)\wedge\de(A-B)$ leads to the gauge-invariant potential $\theta'[A,B,\delta A,\delta B]=\delta(A-B)\wedge(A-B)$, but has equations of motion $\de A=\de B$.} $\de A=0=\de B$. This is therefore a theory of two flat connections, and one can see that by solving half of the equations of motion, i.e. by writing $B=-\de a$, the new fields get pushed to the boundary and \eqref{bulk-extended Abelian CS} becomes \eqref{boundary-extended Lagrangian 1}. Without going on half-shell however, the bulk-extended and boundary-extended Lagrangians differ in a very important way, which has of course to do with the properties of the potential.

The potential for \eqref{bulk-extended Abelian CS} is given by
\be\label{bulk-extended Abelian CS potential}
\theta[A,B,\delta A]=\delta A\wedge(A-B),
\ee
and its gauge transformation is
\be
\theta[\alpha^*A,\alpha^*B,\delta(\alpha^*A)]=\theta[A,B,\delta A]+\de\big(\delta\alpha(A-B)\big)-\delta\alpha\de(A-B).
\ee
Once again, one can make the potential gauge-invariant by adding a field $u$ and considering
\be\label{bulk-extended Abelian CS extended potential}
\theta_{c,p}[A,B,\delta A,\delta u]\coloneqq\theta[A,B,\delta A]+\de\big(\delta u(A-B)\big)-\delta u\de(A-B),
\ee
which satisfies
\be
\theta_{c,p}[\alpha^*A,\alpha^*B,\delta(\alpha^*A),\delta(\alpha^*u)]=\theta_{c,p}[A,B,\delta A,\delta u].
\ee

\subsection{Gauge-invariance and boundary symmetries}

We are now going to compare the properties of the various gauge-invariant extended potentials which we have introduced so far, and in particular study the boundary contribution to their associated pre-symplectic forms.

Before doing so, let us just recall what happens if we work with the gauge-non-invariant potential $\theta[A,\delta A]=\delta A\wedge A$. In this case, the conserved pre-symplectic form only has a bulk piece which is given by
\be
\Omega_\Sigma[A,\delta_1,\delta_2]=-\int_\Sigma\delta A\wedge\delta A,
\ee
and the Hamiltonian generator of the infinitesimal field-independent gauge transformations $\delta_\alpha$ is
\be
\delta\F_c[\alpha]=\Omega_\Sigma[A,\delta,\delta_\alpha]=-2\int_\Sigma\delta A\wedge\de\alpha.
\ee
The reason for which we have denoted this generator by $\F_c[\alpha]$ is that it corresponds actually to the extended generator defined below \eqref{variation of abelian CS flatness constraint}. In particular, it satisfies the properties \eqref{Abelian F_c F_c algebra}. As such, this generator of gauge transformations is not vanishing on-shell, and furthermore possesses an anomalous Poisson algebra. The reason for this is that this extended generator does actually coincide with the boundary observables $\O[\alpha]$.

We are now going to see, following again \cite{Donnelly:2016auv}, how the introduction of the gauge-invariant extended potential enables to describe these boundary observables in terms of a new boundary symmetry, and how it leads to a generator of gauge transformations which is vanishing on-shell.

\subsubsection{Gauge-non-invariant Lagrangian}

The conserved pre-symplectic form originating from the extended potential \eqref{Abelian CS extended potential 1bis} is given by
\be
\Omega_{\Sigma,\partial\Sigma}[A,u,\delta_1,\delta_2]=\Omega_\Sigma[A,\delta_1,\delta_2]+\Omega_{\partial\Sigma}[A,u,\delta_1,\delta_2],
\ee
with
\be
\Omega_\Sigma[A,\delta_1,\delta_2]=-\int_\Sigma\delta A\wedge\delta A,\q\Omega_{\partial\Sigma}[A,u,\delta_1,\delta_2]=-\int_{\partial\Sigma}\delta u\delta(2A+\de u).
\ee
From these expressions, one can get the Hamiltonian generator of the infinitesimal field-independent gauge transformations $\delta_\alpha$. This has now a bulk and a boundary contribution whose sum is given by the variational expression
\be\label{Abelian CS generator with symplectic form}
\slashed{\delta}\F[\alpha]=\slashed{\delta}\F_\Sigma[\alpha]+\slashed{\delta}\F_{\partial\Sigma}[\alpha]=\Omega_\Sigma[A,\delta,\delta_\alpha]+\Omega_{\partial\Sigma}[A,u,\delta,\delta_\alpha]=-2\int_\Sigma\delta A\wedge\de\alpha-2\int_{\partial\Sigma}\alpha\delta A.
\ee
These generators are actually integrable, and satisfy the closed algebra
\be
\lb\F[\alpha],\F[\beta]\rb=\Omega_{\Sigma,\partial\Sigma}[A,u,\delta_\alpha,\delta_\beta]=0
\ee
without the need to restrict the gauge parameters to be vanishing on the boundary. Furthermore, we have that
\be
\F[\alpha]=-2\int_\Sigma\alpha\de A\simeq0.
\ee
By adding the new fields $u$ in the extended potential, the extra boundary piece which contributes to the pre-symplectic form is ensuring that $\F[\alpha]$ is the generator of the infinitesimal transformations $\delta_\alpha$ and that it has a closed algebra, without the need to resort to the discussion below \eqref{variation of abelian CS flatness constraint}. Because of this however, we cannot repeat our previous argument concerning the appearance of the boundary observables $\O[\alpha]$.

These observables are nonetheless still present here, but they are now encoded in a new boundary symmetry. This latter acts on the fields as
\be\label{Abelian CS new boundary symmetry}
\Delta_\alpha A=0,\q\Delta_\alpha u=\alpha,
\ee
and has a generator defined by the variational formula
\be\label{Abelian CS new obervables}
\slashed{\delta}\widetilde{\O}[\alpha]=\Omega_{\partial\Sigma}[A,u,\delta,\Delta_\alpha]=2\int_{\partial\Sigma}\alpha\delta(A+\de u),
\ee
which is integrable if $\delta\alpha=0$. One can see that this is nothing but the previous expression \eqref{Abelian CS corner observable} for the observables, with the difference that the connection is now ``dressed'' by the new boundary field $u$. It is now immediate to see that these symmetry generators satisfy the algebra
\be\label{Abelian CS algebra}
\lb\widetilde{\O}[\alpha],\widetilde{\O}[\beta]\rb=\Omega_{\partial\Sigma}[A,u,\Delta_\alpha,\Delta_\beta]=2\int_{\partial\Sigma}\de\alpha\beta,
\ee
and are observables in the sense that
\be
\lb\F[\alpha],\widetilde{\O}[\beta]\rb=\Omega_{\partial\Sigma}[A,u,\delta_\alpha,\Delta_\beta]=0.
\ee
We have thus described here the boundary symmetry and observables of Abelian Chern--Simons theory using the techniques of \cite{Donnelly:2016auv}. As one can see, the introduction of the fields $u$ in order to obtain a gauge-invariant potential enables to disentangle the notions of gauge transformations and gauge symmetries. The former are generated by a constraint whose generator is vanishing, while the latter are generated by a non-vanishing quantity which is an observable. Furthermore, we see that these observables are given from the onset by a boundary integral, while in \eqref{Abelian CS corner observable} this is only true weakly (i.e. up to a constraint).

\subsubsection{Gauge-invariant boundary-extended Lagrangian}

As explained in \eqref{conserved form with corner fields}, the conserved pre-symplectic form is only sensitive to the presence of a corner term. The pre-symplectic form derived from the three extended gauge-invariant potentials \eqref{Abelian CS extended potential 1bis}, \eqref{Abelian CS extended potential 2}, and \eqref{Abelian CS extended potential 3}, is therefore the same, and the analysis carried out above applies exactly identically to the gauge-invariant boundary-extended Lagrangian.

\subsubsection{Gauge-invariant bulk-extended Lagrangian}

From the extended gauge-invariant potential \eqref{bulk-extended Abelian CS extended potential}, the conserved pre-symplectic form acquires once again a boundary contribution, i.e.
\be
\Omega_{\Sigma,\partial\Sigma}[A,B,u,\delta_1,\delta_2]=\Omega_\Sigma[A,B,\delta_1,\delta_2]+\Omega_{\partial\Sigma}[A,B,u,\delta_1,\delta_2],
\ee
but it is now given by the following bulk and corner pieces:
\be
\Omega_\Sigma[A,B,\delta_1,\delta_2]=-\int_\Sigma\delta A\wedge\delta(A-B),\q\Omega_{\partial\Sigma}[A,B,u,\delta_1,\delta_2]=-\int_{\partial\Sigma}\delta u\delta(A-B).
\ee
From this, we get the generators
\be
\slashed{\delta}\F[\alpha]=-\int_\Sigma\delta(A-B)\wedge\de\alpha-\int_{\partial\Sigma}\alpha\delta(A-B),
\ee
which integrate to
\be
\F[\alpha]=-\int_\Sigma\alpha\de(A-B)\simeq0,
\ee
and satisfy the algebra
\be
\lb\F[\alpha],\F[\beta]\rb=0.
\ee
By plugging \eqref{Abelian CS new boundary symmetry} one gets
\be
\slashed{\delta}\widetilde{\O}[\alpha]=\Omega_{\partial\Sigma}[A,B,u,\delta,\Delta_\alpha]=\int_{\partial\Sigma}\alpha\delta(A-B),
\ee
but these boundary observables turn out to have a vanishing Poisson bracket:
\be
\lb\widetilde{\O}[\alpha],\widetilde{\O}[\beta]\rb=\Omega_{\partial\Sigma}[A,u,\Delta_\alpha,\Delta_\beta]=0.
\ee
Clearly, this is due to the fact that the pre-symplectic form on the boundary is not quadratic in the boundary variable $u$. This illustrate the subtle difference which exists between the Lagrangians \eqref{bulk-extended Abelian CS} and \eqref{boundary-extended Lagrangian 1}. Indeed, although they posses the same equations of motion and the same gauge-invariance property, the former leads to boundary observables with a centrally-extended algebra, while the latter does not. However, one should recall that \eqref{bulk-extended Abelian CS} is defined in the first place with more fields living in the bulk.

\subsection{Noether charges}

Finally, let us end this section by briefly discussing the Noether charges derived from the covariant Hamiltonian analysis. This will illustrate how the corner contribution from the extended potential can be used to obtain gauge-invariant (or vanishing) Noether charges for the gauge symmetries of the theory.

\subsubsection{Gauge-non-invariant Lagrangian}

We first compute the Noether current associated to the infinitesimal transformation generated by $\alpha$. From \eqref{infinitesimal transformation of Abelian CS} one can see that there is a corner ambiguity in this computation since we can write both
\be
\delta_\alpha L[A]=\de(A\wedge\de\alpha)=\de m_1[A,\alpha],\q\delta_\alpha L[A]=\de(\alpha\de A)=\de m_2[A,\alpha].
\ee
These two choices lead to different Noether currents, namely
\be
J_1[A,\alpha]=-2\alpha\de A+2\de(\alpha A),\q J_2[A,\alpha]=-2\alpha\de A+\de(\alpha A),
\ee
and therefore to the two following different conserved Noether charges:
\be
\Q_1[A,\alpha]=\int_{\partial\Sigma}2\alpha A,\q\Q_2[A,\alpha]=\int_{\partial\Sigma}\alpha A.
\ee
One can observe that these Noether charges are not gauge-invariant, i.e. $\delta_\beta\Q_{1,2}[A,\alpha]\neq0$, unless one imposes the condition $\alpha\big|_{\partial\Sigma}=0$.

\subsubsection{Gauge-invariant boundary-extended Lagrangian}

All the conclusions of the previous subsection apply here verbatim. This is due to the fact that, for gauge transformations (i.e. not for diffeomorphisms), the Noether current and the pre-symplectic current are insensitive to the boundary terms $b[\Phi]$ or $\ell[\Phi,\Psi]$ that one can add to the Lagrangian.

\subsubsection{Gauge-invariant bulk-extended Lagrangian}

For the Lagrangian \eqref{bulk-extended Abelian CS}, we have $m[A,B,\alpha]=0$ because of strict gauge-invariance, and the potential is given by \eqref{bulk-extended Abelian CS potential}. This leads to the Noether current
\be
J[A,B,\alpha]=-\alpha\de(A-B)+\de\big(\alpha(A-B)\big),
\ee
from which we can see that the Noether charge
\be
\Q[A,B,\alpha]=\int_{\partial\Sigma}\alpha(A-B)
\ee
is actually gauge-invariant, i.e. $\delta_\beta\Q[A,B,\alpha]=0$.

\subsubsection{Extended pre-symplectic potential}

One can check by an explicit computation that the plugging an infinitesimal gauge transformation into the extended potentials \eqref{Abelian CS extended potential 1bis}, \eqref{Abelian CS extended potential 2}, and \eqref{Abelian CS extended potential 3} leads to a vanishing result, i.e. that
\be
\theta_{c,m,p}[A,u,\delta_\alpha A,\delta_\alpha u]=\theta_{\ell_1,c,m,p}[A,a,u,\delta_\alpha A,\delta_\alpha a,\delta_\alpha u]=\theta_{\ell_2,c,p}[A,a,u,\delta_\alpha A,\delta_\alpha a,\delta_\alpha u]=0.
\ee
This is not surprising, and is simply a consequence of the definition of the extended potential. Now, in the definition \eqref{Noether current} of the Noether current we can choose the corner ambiguity to be the corner contribution appearing in the extended potential.

Explicitly, the infinitesimal gauge-invariance of the extended potential \eqref{Abelian CS extended potential 1} is
\be
\theta_{c,m,p}[A,u,\delta_\alpha A,\delta_\alpha u]=\theta[A,\delta_\alpha A]+\de\big(\delta_\alpha u(2A+\de u)\big)+\delta_\alpha(A\wedge\de u)-2\delta_\alpha u\de A=0.
\ee
Therefore, using the corner term in the definition of the Noether current leads to
\be
J_{1,c}[A,\alpha]
&=\theta_c[A,\delta_\alpha A]-m_1[A,\alpha]\nn\\
&=\theta[A,\delta_\alpha A]+\de\big(\delta_\alpha u(2A+\de u)\big)-A\wedge\de\alpha\nn\\
&=-\delta_\alpha(A\wedge\de u)+2\delta_\alpha u\de A-A\wedge\de\alpha\nn\\
&=-2\alpha\de A+\de(u\de\alpha).
\ee
We see that the charge is not vanishing, but that it is gauge-invariant. The fact that it is non-vanishing is a consequence of the ambiguity in the definition of $m[A,\alpha]$. Indeed, if we consider instead the infinitesimal gauge-invariance of the extended potential \eqref{Abelian CS extended potential 1bis}, we get
\be
\theta_{c,m,p}[A,u,\delta_\alpha A,\delta_\alpha u]=\theta[A,\delta_\alpha A]+\de\big(\delta_\alpha u(A+\de u)-u\delta_\alpha A\big)+\delta_\alpha(u\de A)-2\delta_\alpha u\de A=0.
\ee
Using the corner term in the definition of the Noether current then leads to
\be
J_{2,c}[A,\alpha]
&=\theta_c[A,\delta_\alpha A]-m_2[A,\alpha]\nn\\
&=\theta[A,\delta_\alpha A]+\de\big(\delta_\alpha u(A+\de u)-u\delta_\alpha A\big)-\alpha\de A\nn\\
&=-\delta_\alpha(u\de A)+2\delta_\alpha u\de A-\alpha\de A\nn\\
&=-2\alpha\de A,
\ee
and the charge is therefore vanishing.

This resolves the ambiguity which has been pointed out in the definition of the charges for theories whose Lagrangian is not strictly gauge-invariant. Indeed, even though we cannot use symmetry arguments in order to fix the charge, we can use the corner ambiguity to make it gauge-invariant.

Before moving on to three-dimensional gravity, let us now briefly summarize the results which have been obtained in this section.  We have first reviewed the way in which the boundary dynamics of Abelian Chern--Simons theory is obtained from its Lagrangian and the behavior of gauge transformations. Then, we have recalled how the boundary observables, whose algebra is the current algebra of the boundary WZNW theory, are obtained in the Hamiltonian treatment of \cite{Regge:1974zd,Balachandran:1995qa}. Finally, we have compared these Hamiltonian boundary observables with the observables arising from the extended phase space treatment of \cite{Donnelly:2016auv}

\section{First order gravity}
\label{sec:4}

We now turn to the case of three-dimensional gravity in its first order formulation. We are first going to recall the form of the infinitesimal and finite gauge transformations. Following what we did for Chern--Simons theory, we will then recall the structure of the Hamiltonian analysis in the presence of boundaries and describe how boundary observables arise in this framework. We will then apply the framework of \cite{Donnelly:2016auv} to (one choice of parametrization of) the gauge transformations of first-order three-dimensional gravity, and obtain the extended potential and symplectic structure which lead to the boundary symmetries and observables.

\subsection{Lagrangian}

The Lagrangian for (Euclidean) first order gravity with a cosmological constant is
\be\label{first order gravity Lagrangian}
L[e,\omega]\coloneqq\tr\left(e\wedge F+\f{1}{6\ell^2}e\wedge[e\wedge e]\right).
\ee
Its variation is given by
\be\label{variation of first order gravity Lagrangian}
\delta L[e,\omega]=\tr\left(\delta e\wedge\left(F+\f{1}{2\ell^2}[e\wedge e]\right)+\delta\omega\wedge\De e\right)+\de\,\tr(\delta\omega\wedge e),
\ee
from which we can read the pre-symplectic potential.

The symmetries of the theory are the $\SU(2)$ gauge transformations, the so-called translations, and the diffeomorphisms. The action of infinitesimal $\SU(2)$ gauge transformations is given by
\be\label{first order infinitesimal gauge transformation}
\delta_\alpha^\text{g}e=[e,\alpha],\q\delta_\alpha^\text{g}\omega=\De\alpha,\q\delta_\alpha^\text{g}F=[F,\alpha],
\ee
where $\alpha\in\Omega^0\big(M,\su(2)\big)$. When acting on the Lagrangian, this is of course
\be
\delta_\alpha^\text{g}L[e,\omega]=0.
\ee

For the infinitesimal translations, the transformation of the various fields is given by
\be\label{first order infinitesimal translation}
\delta_\phi^\text{t}e=\De\phi,\q\delta_\phi^\text{t}\omega=\f{1}{\ell^2}[e,\phi],\q\delta_\phi^\text{t}F=\f{1}{\ell^2}\De[e,\phi],
\ee
where $\phi\in\Omega^0\big(M,\su(2)\big)$ is an infinitesimal generator. At the level of the Lagrangian, we have
\be\label{infinitesimal topological translation of first order action}
\delta_\phi^\text{t}L[e,\omega]=\de\,\tr\left(\phi\left(F+\f{1}{2\ell^2}[e\wedge e]\right)-\f{1}{\ell^2}\phi[e\wedge e]\right),
\ee
showing that if the cosmological constant is not vanishing the Lagrangian is not on-shell gauge-invariant. Note that this on-shell non-invariance of the Lagrangian under certain infinitesimal gauge transformations also appears in the case of non-Abelian Chern--Simons theory, as can be seen on \eqref{infinitesimal transfo of non-Abelian CS}, and for diffeomorphisms in metric general relativity with a non-vanishing cosmological constant. It can actually be traced back to the fact that non-Abelian Chern--Simons theory and gravity with a non-zero cosmological constant are defined by Lagrangians which are not vanishing on-shell. Since in this paper we will construct the extended phase space of first order gravity in the case of a vanishing cosmological constant, we will not be bothered by this fact. However, when extending these results and that of \cite{Donnelly:2016auv} to the case of a non-zero cosmological constant, additional subtleties have to be handled and are discussed in \cite{Speranza:2017gxd,MGtoappear}.

Now, from \eqref{first order infinitesimal gauge transformation} and \eqref{first order infinitesimal translation}, we can compute the algebra structure
\be\label{infinitesimal algebra}
[\delta_\alpha^\text{g},\delta_\beta^\text{g}]=\delta_{[\alpha,\beta]}^\text{g},\q[\delta_\phi^\text{t},\delta_\chi^\text{t}]=\f{1}{\ell^2}\delta_{[\phi,\chi]}^\text{g},\q[\delta_\alpha^\text{g},\delta_\phi^\text{t}]=\delta_{[\alpha,\phi]}^\text{t},
\ee
which is nothing but that of $\so(4)$. We will compare this later on with the Poisson algebra of the generators of these infinitesimal transformations. As is well-known, these two algebras will turn out to be equal up to a central extension.

Finally, the infinitesimal action of diffeomorphisms is parametrized by a vector field $\xi$ and given by the Lie derivative $\L_\xi=\de(\xi\ip\cdot)+\xi\ip(\de\cdot)$. Explicitly, this is
\be
&\delta_\xi^\text{d}e_\mu=(\L_\xi e)_\mu=\big(\de(\xi\ip e)+\xi\ip(\de e)\big)_\mu=\partial_\mu\xi^\nu e_\nu+\xi^\nu\partial_\nu e_\mu,\nn\\
&\delta_\xi^\text{d}\omega_\mu=(\L_\xi\omega)_\mu=\big(\de(\xi\ip\omega)+\xi\ip(\de\omega)\big)_\mu=\partial_\mu\xi^\nu\omega_\nu+\xi^\nu\partial_\nu\omega^i_\mu.
\ee
One can then check that we have the well-known relations
\be\label{relation d t g for e}
\delta_{\xi\ip\omega}^\text{g}e+\delta_{\xi\ip e}^\text{t}e+\xi\ip(\De e)
&=[e,\xi\ip\omega]+\De(\xi\ip e)+\xi\ip(\De e)\nn\\
&=[e,\xi\ip\omega]+\de(\xi\ip e)+[\omega,\xi\ip e]+\xi\ip(\de e)+\xi\ip[\omega,e]\nn\\
&=\de(\xi\ip e)+\xi\ip(\de e)\nn\\
&=\L_\xi e\nn\\
&=\delta_\xi^\text{d}e,
\ee
and
\be\label{relation d t g for omega}
\delta_{\xi\ip\omega}^\text{g}\omega+\delta_{\xi\ip e}^\text{t}\omega+\xi\ip\left(F+\f{1}{2\ell^2}[e\wedge e]\right)
&=\De(\xi\ip\omega)+\f{1}{\ell^2}[e,\xi\ip e]+\xi\ip\left(F+\f{1}{2\ell^2}[e\wedge e]\right)\nn\\
&=\de(\xi\ip\omega)+[\omega,\xi\ip\omega]+\xi\ip(\de\omega)+\f{1}{2}\xi\ip[\omega\wedge\omega]\nn\\
&=\de(\xi\ip\omega)+\xi\ip(\de\omega)\nn\\
&=\L_\xi\omega\nn\\
&=\delta_\xi^\text{d}\omega.
\ee
This means that, on-shell, the action of diffeomorphisms can be written as a combination of field-dependent gauge transformations and translations. In other words, we have that
\be\label{diffeo gauge and top}
\delta_\xi^\text{d}\simeq\delta_{\xi\ip\omega}^\text{g}+\delta_{\xi\ip e}^\text{t}.
\ee
We can now discuss the finite action of these gauge transformations.

Under finite $\SU(2)$ gauge transformations parametrized by a group element $h$, the triad, the connection, and its curvature, transform as
\be\label{finite SU(2) gauge transformations}
h^*e=h^{-1}eh,\q h^*\omega=h^{-1}\omega h+h^{-1}\de h,\q h^*F=h^{-1}Fh.
\ee
At the level of the Lagrangian, this implies that
\be
L[h^*e,h^*\omega]=L[e,\omega].
\ee

The finite version of the translations takes a different form depending on whether there is a cosmological constant or not. If the cosmological constant is vanishing, i.e. for $\ell^2=\infty$, we have
\be\label{finite translation e omega}
\phi^*e=e+\De\phi,\q\phi^*\omega=\omega\q\phi^*F=F,
\ee
where $\phi\in\Omega^0\big(M,\su(2)\big)$ is now a finite parameter instead of an infinitesimal generator. At the level of the Lagrangian this leads to
\be\label{finite topological transformation of gravity}
L[\phi^*e,\phi^*\omega]=L[e,\omega]+\de\,\tr(\phi F)\simeq L[e,\omega],
\ee
so the Lagrangian is invariant on-shell under finite translations. In the case of a non-vanishing cosmological constant on the other hand, the finite transformations are parametrized by a group element $t\in\SU(2)$. Their action on the triad is given by
\be\label{finite translation e lambda}
t^*e
&=\f{\ell}{2}\left[t^{-1}\left(\omega+\f{1}{\ell}e\right)t-t\left(\omega-\f{1}{\ell}e\right)t^{-1}+t^{-1}\de t-t\de t^{-1}\right]\nn\\
&=\f{\ell}{2}\left[t_\text{g}^*\left(\omega+\f{1}{\ell}e\right)-(t_\text{g}^{-1})^*\left(\omega-\f{1}{\ell}e\right)\right],
\ee
while their action on the connection is given by
\be\label{finite translation omega lambda}
t^*\omega
&=\f{1}{2}\left[t^{-1}\left(\omega+\f{1}{\ell}e\right)t+t\left(\omega-\f{1}{\ell}e\right)t^{-1}+t^{-1}\de t+t\de t^{-1}\right]\nn\\
&=\f{1}{2}\left[t_\text{g}^*\left(\omega+\f{1}{\ell}e\right)+(t_\text{g}^{-1})^*\left(\omega-\f{1}{\ell}e\right)\right].
\ee
Here we have simply used $t_\text{g}^*$ to denote the action of $t$ not as a translation but as a gauge transformation of the type \eqref{finite SU(2) gauge transformations}. This shows clearly that $t^*(e,\omega)\big|_{t=\openone}=(e,\omega)$ and that at the infinitesimal level $t^*$ becomes indeed $\delta^\text{t}_{\ell\phi}$ (notice that $\phi$ picks up here a dimensional factor $\ell$).

Finally, for a diffeomorphism $Y:M\rightarrow M$, the finite action is given by the pullback maps
\be
e_\mu(x)\mapsto(Y^*e)_\mu(x)=\partial_\mu Y^\nu(x)e_\nu\big(Y(x)\big),\q\omega_\mu(x)\mapsto(Y^*\omega)_\mu(x)=\partial_\mu Y^\nu(x)\omega_\nu\big(Y(x)\big).
\ee

\subsection{Hamiltonian}

In this part we are going to study the extended Hamiltonian generators and their algebra. First of all, the Hamiltonian action takes the form
\be\label{Hamiltonian first order action}
S[e,\omega]
&=\f{1}{2}\int_M\de^3x\,\teps^{\mu\nu\rho}\tr\left(e_\mu F_{\nu\rho}+\f{1}{3\ell^2}e_\mu[e_\nu,e_\rho]\right)\nn\\
&=\f{1}{2}\int_{\mathbb{R}\times\Sigma}\de t\,\de^2x\,\teps^{ab}\tr\left(e_0F_{ab}+2e_aF_{b0}+\f{1}{\ell^2}e_0[e_a,e_b]\right)\nn\\
&=\int_\mathbb{R}\de t\int_\Sigma\tr\left(\partial_0\omega\wedge e+\omega_0\De e+e_0\left(F+\f{1}{2\ell^2}[e\wedge e]\right)-\de(e\omega_0)\right),
\ee
with canonical Poisson bracket $\lb e_a(x),\omega_b(y)\rb=-\teps_{ab}\delta^2(x,y)$, and generic brackets
\be
\lb f_1,f_2\rb=-\teps_{ab}\int_\Sigma\de^2x\int_\Sigma\de^2y\,\delta^2(x,y)\left(\f{\delta f_1}{\delta e_a(x)}\f{\delta f_2}{\delta\omega_b(y)}-\f{\delta f_2}{\delta e_a(x)}\f{\delta f_1}{\delta\omega_b(y)}\right).
\ee
Here the multipliers $\omega_0$ and $e_0$ are enforcing respectively the Gauss and flatness constraints corresponding to the infinitesimal $\SU(2)$ gauge transformations and to the translations. These are the gauge transformations which we will focus on for the study of the pre-symplectic potential and of the boundary observables.

Let us now replace the multipliers $\omega_0$ and $e_0$ by arbitrary smearing parameters and introduce the smeared Gauss and curvature constraints
\be
\G[\alpha]\coloneqq-\int_\Sigma\tr(\alpha\De e),\q\F[\phi]\coloneqq-\int_\Sigma\tr\left(\phi\left(F+\f{1}{2\ell^2}[e\wedge e]\right)\right).
\ee
The variations are then given by
\be
\delta\G[\alpha]=-\int_\Sigma\tr(\delta\alpha\De e+\delta e\wedge\De\alpha+\delta\omega\wedge[e,\alpha])-\int_{\partial\Sigma}\tr(\alpha\delta e),
\ee
and
\be
\delta\F[\phi]=-\int_\Sigma\tr\left(\delta\phi\left(F+\f{1}{2\ell^2}[e\wedge e]\right)+\delta\omega\wedge\De\phi+\f{1}{\ell^2}\delta e\wedge[e,\phi]\right)-\int_{\partial\Sigma}\tr(\phi\delta\omega).
\ee
In order to obtain functionally differentiable quantities, we can then consider the extended generators defined by
\be
\slashed{\delta}\G_c[\alpha]\coloneqq\delta\G[\alpha]+\int_{\partial\Sigma}\tr(\alpha\delta e),\q\slashed{\delta}\F_c[\phi]\coloneqq\delta\F[\phi]+\int_{\partial\Sigma}\tr(\phi\delta\omega).
\ee
For $\delta\alpha=0=\delta\phi$, these extended generators are integrable and generate the infinitesimal gauge transformations \eqref{first order infinitesimal gauge transformation} and \eqref{first order infinitesimal translation}. Furthermore, they satisfy an algebra which has a first class part
\be
\lb\G_c[\alpha],\G_c[\beta]\rb=\G_c\big[[\alpha,\beta]\big],\q\lb\F_c[\phi],\F_c[\varphi]\rb=\f{1}{\ell^2}\G_c\big[[\phi,\varphi]\big],
\ee
and an anomalous part
\be
\lb\G_c[\alpha],\F_c[\phi]\rb=\F_c\big[[\alpha,\phi]\big]+\int_{\partial\Sigma}\tr(\de\alpha\phi).
\ee
As expected, we see here that the Poisson algebra of the extended generators differs from the algebra structure \eqref{infinitesimal algebra} by the presence of a central extension.

In order for this last bracket to close, one has to consider generators with compact support, i.e. $\bar{\alpha}\big|_{\partial\Sigma}=0=\bar{\phi}\big|_{\partial\Sigma}$. In this case, boundary observables arise following the same mechanism as in sections \ref{subsec:Hamiltonian Abelian CS} and \ref{subsec:Hamiltonian CS}. In order to describe them, let us focus on the case $\ell^2=\infty$. For the $\su(2)$ transformations and the translations, we have respectively the observables
\be\label{first order boundary observables 1}
\O^\text{g}[\alpha]\coloneqq-\int_\Sigma\tr(e\wedge\De\alpha)=\G[\alpha]+\int_{\partial\Sigma}\tr(\alpha e)\simeq\int_{\partial\Sigma}\tr(\alpha e),
\ee
and
\be\label{first order boundary observables 2}
\O^\text{t}[\phi]\coloneqq-\int_\Sigma\tr\left(\omega\wedge\de\phi+\f{1}{2}\phi[\omega\wedge\omega]\right)=\F[\phi]+\int_{\partial\Sigma}\tr(\phi\omega)\simeq\int_{\partial\Sigma}\tr(\phi\omega),
\ee
and their algebra is given by
\be\label{first order boundary observables algebra}
\lb\O^\text{g}[\alpha],\O^\text{t}[\phi]\rb=\O^\text{t}\big[[\alpha,\phi]\big]+\int_{\partial\Sigma}\tr(\de\alpha\phi).
\ee

We are now going to analyse the behavior of the potential under finite field-dependent $\SU(2)$ gauge transformations and translations. As we will see, an extra compatibility condition will be required in order to obtain an extended potential which is gauge-invariant under both types of transformations. From this, we will then be able to proceed with the analysis of the generators of gauge transformations and boundary symmetries. As we have seen in the case of Abelian Chern--Simons theory, the generators of the boundary symmetries will be the boundary observables of the theory.

\subsection[$\SU(2)$ gauge transformations]{$\boldsymbol{\SU(2)}$ gauge transformations}

Recall that the potential of the Lagrangian $L[e,\omega]$ is given by
\be
\theta[e,\delta\omega]=\tr(\delta\omega\wedge e).
\ee
Since the Lagrangian is strictly gauge-invariant, i.e. $L[h^*e,h^*\omega]=L[e,\omega]$, we have that
\be
\delta(L[h^*e,h^*\omega])=\delta L[e,\omega]=E[\Phi]\wedge\delta\Phi+\de\theta[e,\delta\omega],
\ee
where the term giving the equations of motion is simply the bulk term appearing in \eqref{variation of first order gravity Lagrangian}. Now, using the identity $\delta(h^{-1}\de h)=h^{-1}\de(\delta hh^{-1})h$ as well as the two very important relations\footnote{Notice that here it is crucial for our purposes to keep terms of the form $\delta h$, i.e. to consider that $h$ are parameters of field-dependent gauge transformations.}
\be\label{variation of transformation of omega and e}
\delta(h^*\omega)=h^{-1}\big(\delta\omega+\De(\delta hh^{-1})\big)h,\q\delta(h^*e)=h^{-1}(\delta e+[e,\delta hh^{-1}])h,
\ee
we get
\be\label{finite SU(2) transformation of EOM term}
(\delta L)[h^*e,h^*\omega]=E[\Phi]\wedge\delta\Phi+\de\,\tr\big(e\wedge\De(\delta hh^{-1})\big)+\de\theta[h^*e,\delta(h^*\omega)].
\ee
The proof of this result is given in appendix \ref{appendix:proofs}. This is not yet of the form \eqref{transformation of variation of Lagrangian}, but using the fact that
\be\label{IBP of eDh}
e\wedge\De(\delta hh^{-1})=\De e\delta hh^{-1}-\de(e\delta hh^{-1})
\ee
does lead to the expected form of the transformation, namely to the total derivative of a term involving the equation of motion $\De e\simeq0$.

We can now clearly anticipate what will happen when computing the gauge transformation of the potential. Using again \eqref{variation of transformation of omega and e} and \eqref{IBP of eDh}, one gets that
\be\label{SU(2) transformation of bare potential}
\theta[h^*e,\delta(h^*\omega)]
&=\theta[e,\delta\omega]+\tr\big(\De(\delta hh^{-1})\wedge e\big)\nn\\
&=\theta[e,\delta\omega]+\de\,\tr(e\delta hh^{-1})-\tr(\De e\delta hh^{-1}).
\ee
What we see here is simply the manifestation of the fact that \cite{Donnelly:2016auv,Gomes:2016mwl}
\be\label{commuting delta and SU(2)}
\delta(h^*\omega)=h^{-1}\big(\delta\omega+\De(\delta hh^{-1})\big)h=h^*\big(\delta\omega+\De(\delta hh^{-1})\big)=h^*\big(\delta\omega+\delta^\text{g}_{\delta hh^{-1}}\omega\big),
\ee
where we have to remember that $h^*(\delta\omega)=h\delta\omega h^{-1}$ since $\delta\omega$ is the difference between two connections. Similarly, we have that
\be\label{commuting delta and top}
\delta(h^*e)=h^{-1}(\delta e+[e,\delta hh^{-1}])h=h^*(\delta e+[e,\delta hh^{-1}])=h^*\big(\delta e+\delta^\text{g}_{\delta hh^{-1}}e\big).
\ee
Now, we can make the potential gauge-invariant on-shell by adding a field $u\in\Omega^0\big(\partial\Sigma,\SU(2)\big)$, which is a choice of local trivialization transforming as $h^*u=h^{-1}u$. With this, one can check that
\be\label{composition of h and u}
\delta(h^*u)(h^*u)^{-1}=h^{-1}(\delta uu^{-1}-\delta hh^{-1})h=h^*(\delta uu^{-1}-\delta hh^{-1}).
\ee
Defining the extended potential
\be\label{first order SU(2) extended potential}
\theta_{c,p}[e,\omega,u,\delta\omega,\delta u]\coloneqq\theta[e,\delta\omega]+\de\,\tr(e\delta uu^{-1})-\tr(\De e\delta uu^{-1}),
\ee
one can see that
\be
\theta_{c,p}[h^*e,h^*\omega,h^*u,\delta(h^*\omega),\delta(h^*u)]=\theta_{c,p}[e,\omega,u,\delta\omega,\delta u].
\ee
This defines an extended potential which is gauge-invariant under field-dependent $\SU(2)$ gauge transformations. We now turn to the case of the translations.

\subsection{Translations}

As one can see by comparing \eqref{finite translation e omega} with \eqref{finite translation e lambda} and \eqref{finite translation omega lambda}, the finite translations have a drastically different structure depending on whether the cosmological constant is vanishing or not. For simplicity we will focus here on the case $\ell^2=\infty$. In addition, we have seen in \eqref{finite topological transformation of gravity} that the Lagrangian is not gauge-invariant under the translations, but transforms with a boundary term. This is analogous to what happened with the gauge transformations in Abelian Chern--Simons theory. For gravity, it is also possible to define Lagrangians which are gauge-invariant from the onset, by adding either a boundary or a bulk field. We discuss this possibility in appendix \ref{appendix:top}.

Let us now focus on the Lagrangian $L[e,\omega]$ with $\ell^2=\infty$. Computing the variation of \eqref{finite topological transformation of gravity}, we get that
\be
\delta(L[\phi^*e,\phi^*\omega])=E[\Phi]\wedge\delta\Phi+\de(\theta[e,\delta\omega]+\phi F).
\ee
On the other hand, the gauge transformation of the variation of the Lagrangian leads to
\be\label{finite topological translation transformation of EOM term}
(\delta L)[\phi^*e,\phi^*\omega]=E[\Phi]\wedge\delta\Phi+\de\,\tr(\delta\phi F)+\de\theta[\phi^*e,\delta(\phi^*\omega)].
\ee
This is shown in appendix \ref{appendix:proofs}, and, importantly, does only hold in the case $\ell^2=\infty$. Once again, one can compute the gauge transformation of the potential to find
\be\label{topological translation of bare potential}
\theta[\phi^*e,\delta(\phi^*\omega)]
&=\theta[e,\delta\omega]+\tr(\delta\omega\wedge\De\phi)\nn\\
&=\theta[e,\delta\omega]-\de\,\tr(\phi\delta\omega)+\tr(\phi\delta F)\nn\\
&=\theta[e,\delta\omega]-\de\,\tr(\phi\delta\omega)+\delta\,\tr(\phi F)-\tr(\delta\phi F),
\ee
which is exactly of the form \eqref{transformation of potential}, and shows that $\delta(L[\phi^*e,\phi^*\omega])=(\delta L)[\phi^*e,\phi^*\omega]$.

Now, we can make the potential gauge-invariant by adding a field $\chi\in\Omega^0\big(\partial\Sigma,\su(2)\big)$ which transforms as $\phi^*\chi=\chi-\phi$. Defining the extended potential
\be\label{first order topological extended potential}
\theta_{c,m,p}[e,\omega,\chi,\delta\omega]\coloneqq\theta[e,\delta\omega]-\de\,\tr(\chi\delta\omega)+\delta\,\tr(\chi F)-\tr(\delta\chi F),
\ee
one can then see that
\be
\theta_{c,m,p}[\phi^*e,\phi^*\omega,\phi^*\chi,\delta(\phi^*\omega)]\simeq\theta_{c,m,p}[e,\omega,\chi,\delta\omega].
\ee

Before moving on to the study of the compatibility between the extended gauge-invariant potentials \eqref{first order SU(2) extended potential} and \eqref{first order topological extended potential}, let us briefly illustrate once again how the corner ambiguities can be used to obtain a vanishing Noether charge for the gauge transformations.

\subsection{Noether charges}

For infinitesimal $\SU(2)$ gauge transformations, we have $\delta_\alpha^\text{g} L[e,\omega]=0$. This implies that the associated Noether current is
\be
J[e,\omega,\alpha]=\theta[e,\delta_\alpha^\text{g}\omega]=-\tr(\alpha\De e)+\de\,\tr(\alpha e).
\ee
Now, the extended potential \eqref{first order SU(2) extended potential} is such that
\be
\theta_{c,p}[e,\omega,u,\delta_\alpha^\text{g}\omega,\delta_\alpha^\text{g}u]=\theta[e,\delta_\alpha^\text{g}\omega]+\de\,\tr(e\delta_\alpha^\text{g}uu^{-1})-\tr(\De e\delta_\alpha^\text{g} uu^{-1})=0.
\ee
Therefore, using the corner term in the definition of the Noether current leads to
\be
J_c[e,\omega,\alpha]=\theta_c[e,\omega,u,\delta_\alpha^\text{g}\omega,\delta_\alpha^\text{g}u]=\theta[e,\delta_\alpha^\text{g}\omega]+\de\,\tr(e\delta_\alpha^\text{g}uu^{-1})=\tr(\De e\delta_\alpha^\text{g}uu^{-1})=-\tr(\alpha\De e),
\ee
thereby showing that the charge is indeed vanishing.

For infinitesimal translations, the initial Lagrangian transforms as $\delta_\phi^\text{t}L[e,\omega]=\de\,\tr(\phi F)$. This implies that the associated Noether current is
\be
J[e,\omega,\phi]=\theta[e,\delta_\phi^\text{t}\omega]-m[\omega,\phi]=-\tr(\phi F),
\ee
so the charge is vanishing from the onset. This is of course consistent with the extended potential, since one can see from \eqref{first order topological extended potential} that $\theta_{c,m,p}[e,\omega,\chi,\delta_\phi^\text{t}\omega]$ is actually vanishing term-by-term.

\subsection[Compatibility between $\SU(2)$ and translations]{Compatibility between $\boldsymbol{\SU(2)}$ and translations}
\label{subsec:compatibility}

In order to obtain gauge-invariance of the potential under both the $\SU(2)$ gauge transformations and the translations (which we recall are studied in the case $\ell^2=\infty$), we have to somehow combine the extended potentials. For this, we need to know how the transformations act on the new boundary fields. This is given by
\be
h^*\chi=h^{-1}\chi h,\q\delta(\phi^*u)(\phi^*u)^{-1}=\delta uu^{-1}.
\ee
With this, we can compute the action of translations on \eqref{first order SU(2) extended potential}, which is
\be
\theta_{c,p}[\phi^*e,\phi^*\omega,\phi^*u,\delta(\phi^*\omega),\delta(\phi^*u)]=\theta_{c,p}[e,\omega,u,\delta\omega,\delta u]+\tr(\delta\omega\wedge\De\phi)+\tr\big(\De(\delta uu^{-1})\wedge\De\phi\big).
\ee
On the other hand, the action of $\SU(2)$ gauge transformations on \eqref{first order topological extended potential} is given by
\be
\theta_{c,m,p}[h^*e,h^*\omega,h^*\chi,\delta(h^*\omega)]=\theta_{c,m,p}[e,\omega,\chi,\delta\omega]+\tr\big(\De(\delta hh^{-1})\wedge e\big)+\tr\big(\De(\delta hh^{-1})\wedge\De\chi\big).
\ee
From these two equations, it is clear that a new term is needed in order to obtain a potential which is invariant under both transformations.

The fully-invariant potential is actually given by
\be\label{fully-invariant first order potential on-shell}
\theta[e,\omega,u,\chi,\delta\omega,\delta u]
&\coloneqq\theta[e,\delta\omega]+\tr\big(\De(\delta uu^{-1})\wedge e\big)+\tr(\delta\omega\wedge\De\chi)+\tr\big(\De(\delta uu^{-1})\wedge\De\chi\big)\nn\\
&=\theta[e,\delta\omega]+\de\,\tr\big((e+\De\chi)\delta uu^{-1}-\chi\delta\omega\big)-\tr\big((\De e+[F,\chi])\delta uu^{-1}-\chi\delta F\big)\nn\\
&\simeq\theta[e,\delta\omega]+\de\,\tr\big((e+\De\chi)\delta uu^{-1}-\chi\delta\omega\big),
\ee
as one can easily check. In the second and third lines, we have simply isolated the equations of motion and then gone on-shell to obtain the expression which is suitable for the computation of the conserved extended pre-symplectic form. We now have all the ingredients necessary for the computation of the generators of gauge transformations and boundary symmetries.

Let us simply observe at this point that this extended potential can also be written in the suggestive form
\be\label{dressed potential}
\theta[e,\omega,u,\chi,\delta\omega,\delta u]=\tr(\delta\tilde{\omega}\wedge\tilde{e}),
\ee
where we have introduced the ``dressed'' fields $\tilde{e}\coloneqq u^*(e+\De\chi)=u^*(\chi^*e)$ and $\tilde{\omega}\coloneqq u^*\omega$. Furthermore, one can check that this expression is of course compatible with the usual correspondance between Chern--Simons theory (which would be here for the gauge group ISU(2)) and three-dimensional gravity. Indeed, just like the potential $\la\delta A\wedge A\ra$ for Chern--Simons theory reduces to $\tr(\delta\omega\wedge e)$ (with the standard non-trivial paring between rotations and translations) once the connection $A$ is decomposed in terms of $e$ and $\omega$, one could check using the results of \ref{appendix:CS} that the extended potential of Chern--Simons theory reduces as expected to the extended potential \eqref{fully-invariant first order potential on-shell} if the new Chern--Simons boundary field $g\in\text{ISU}(2)$ is written with the Cartan decomposition $g=u\chi$.

\subsection{Gauge-invariance and boundary symmetries}

We are now once again going to follow \cite{Donnelly:2016auv} in order to construct the symplectic form, and extract from it the generators of gauge transformations and the boundary observables generating the boundary symmetries.

To compute the conserved pre-symplectic form arising from the potential \eqref{fully-invariant first order potential on-shell}, we have to use the identity
\be
\delta(\delta uu^{-1})=-\delta u\delta u^{-1}=-\delta uu^{-1}u\delta u^{-1}=\delta uu^{-1}\delta uu^{-1}=\f{1}{2}[\delta uu^{-1},\delta uu^{-1}].
\ee
With this, we then find
\be
\Omega_{\Sigma,\partial\Sigma}[e,\omega,u,\chi,\delta_1,\delta_2]=\Omega_\Sigma[e,\omega,\delta_1,\delta_2]+\Omega_{\partial\Sigma}[e,\omega,u,\chi,\delta_1,\delta_2],
\ee
with the bulk pre-symplectic form
\be
\Omega_\Sigma[e,\omega,\delta_1,\delta_2]=-\int_\Sigma\tr(\delta\omega\wedge\delta e),
\ee
and the boundary contribution
\be
\Omega_{\partial\Sigma}[e,\omega,u,\chi,\delta_1,\delta_2]=\int_{\partial\Sigma}\tr\left(\delta(e+\De\chi)\delta uu^{-1}+\f{1}{2}(e+\De\chi)[\delta uu^{-1},\delta uu^{-1}]-\delta\chi\delta\omega\right).
\ee
From these expressions for the bulk and the boundary contributions to the conserved pre-symplectic form, we are going to derive the expressions for the generators of gauge transformations and for the observables generating boundary symmetries.

In order to derive the expression for the generators of the infinitesimal gauge transformations, we need to know how the boundary fields $u$ and $\chi$ transform under $\delta_\alpha^\text{g}$ and $\delta_\phi^\text{t}$. This is given by
\be
\delta_\alpha^\text{g}u=-\alpha u,\q\delta_\alpha^\text{g}\chi=[\chi,\alpha],\q\delta_\phi^\text{t}u=0,\q\delta_\phi^\text{t}\chi=-\phi.
\ee
We therefore find that the generator of infinitesimal $\SU(2)$ gauge transformations is given by
\be\label{new SU(2) generator}
\slashed{\delta}\G[\alpha]=\Omega_\Sigma[e,\omega,\delta,\delta_\alpha^\text{g}]+\Omega_{\partial\Sigma}[e,\omega,u,\chi,\delta,\delta_\alpha^\text{g}]=-\int_\Sigma\tr(\delta e\wedge\De\alpha+\delta\omega\wedge[e,\alpha])-\int_{\partial\Sigma}\tr(\alpha\delta e),
\ee
while the generator of infinitesimal translations is given by
\be\label{new translation generator}
\slashed{\delta}\F[\phi]=\Omega_\Sigma[e,\omega,\delta,\delta_\phi^\text{t}]+\Omega_{\partial\Sigma}[e,\omega,u,\chi,\delta,\delta_\phi^\text{t}]=-\int_\Sigma\tr(\delta\omega\wedge\De\phi)-\int_{\partial\Sigma}\tr(\phi\delta\omega).
\ee
We can now compute the algebra between these generators to find
\be
\lb\G[\alpha],\G[\beta]\rb=\delta_\alpha^\text{g}\G[\beta]=\G\big[[\alpha,\beta]\big],\q\lb\F[\phi],\F[\varphi]\rb=\delta_\phi^\text{t}\F[\varphi]=0,
\ee
and
\be
\lb\G[\alpha],\F[\phi]\rb=\delta_\alpha^\text{g}\F[\phi]=\F\big[[\alpha,\phi]\big],
\ee
which is not anomalous anymore, and closes without the need to restrict the gauge parameters on $\partial\Sigma$. This is to be put in parallel with the results obtained previously with the extended generators.

Our task is now to find a description of the boundary symmetries and of the observables generating them. Since we have constructed our pre-symplectic form by studying the $\SU(2)$ gauge transformations and the translations, we naturally expect to find two boundary symmetries, as well as two types of boundary observables generalizing \eqref{first order boundary observables 1} and \eqref{first order boundary observables 2}. First there exists a boundary symmetry which acts on the fields as\footnote{Recall that $u$ is here a group element, while in \eqref{Abelian CS new boundary symmetry} it is a Lie algebra element.}
\be\label{SU(2) boundary symmetry}
\Delta_\alpha^\text{g}e=0,\q\Delta_\alpha^\text{g}\omega=0,\q\Delta_\alpha^\text{g} u=u\alpha,\q\Delta_\alpha^\text{g}\chi=0,
\ee
and whose generator is given by
\be
\slashed{\delta}\widetilde{\O}^\text{g}[\alpha]=\Omega_{\partial\Sigma}[e,\omega,u,\chi,\delta,\Delta_\alpha^\text{g}]=\int_{\partial\Sigma}\tr\Big(u\alpha u^{-1}\big(\delta(e+\De\chi)+[e+\De\chi,\delta uu^{-1}]\big)\Big).
\ee
Notice that this can be rewritten in terms of the dressed fields $\tilde{e}=u^*(e+\De\chi)=u^*(\chi^*e)$ (this is well-defined since the finite actions $u^*$ and $\chi^*$ commute) as
\be\label{new boundary observable SU(2)}
\slashed{\delta}\widetilde{\O}^\text{g}[\alpha]=\int_{\partial\Sigma}\tr(\alpha\delta\tilde{e}),
\ee
and it is then explicit that the expression $\slashed{\delta}\widetilde{\O}^\text{g}[\alpha]$ is integrable. The generators satisfy the algebra
\be
\lb\widetilde{\O}^\text{g}[\alpha],\widetilde{\O}^\text{g}[\beta]\rb=\Omega_{\partial\Sigma}[e,\omega,u,\chi,\Delta_\alpha^\text{g},\Delta_\beta^\text{g}]=\widetilde{\O}^\text{g}\big[[\alpha,\beta]\big],
\ee
and are observables in the sense that
\be
\lb\G[\alpha],\widetilde{\O}^\text{g}[\beta]\rb=\Omega_{\partial\Sigma}[e,\omega,u,\chi,\delta_\alpha^\text{g},\Delta_\beta^\text{g}]=0,\q\lb\F[\phi],\widetilde{\O}^\text{g}[\beta]\rb=\Omega_{\partial\Sigma}[e,\omega,u,\chi,\delta_\phi^\text{t},\Delta_\beta^\text{g}]=0.
\ee
We have therefore obtained a new observable $\widetilde{\O}^\text{g}[\alpha]$ which is a ``dressed'' version of \eqref{first order boundary observables 1} under the action of both boundary fields $u$ and $\chi$.

Now, there exists also a boundary symmetry inherited from the translations. This symmetry acts on the fields as
\be\label{topological boundary symmetry}
\Delta_\phi^\text{t}e=0,\q\Delta_\phi^\text{t}\omega=0,\q\Delta_\phi^\text{t}u=0,\q\Delta_\phi^\text{t}\chi=u\phi u^{-1},
\ee
and is generated by
\be
\slashed{\delta}\widetilde{\O}^\text{t}[\phi]=\Omega_{\partial\Sigma}[e,\omega,u,\chi,\delta,\Delta_\phi^\text{t}]=\int_{\partial\Sigma}\tr\Big(u\phi u^{-1}\big(\delta\omega+\De(\delta uu^{-1})\big)\Big).
\ee
This can be rewritten in terms of the dressed field $\tilde{\omega}=u^*\omega$ as
\be\label{new boundary observable top}
\slashed{\delta}\widetilde{\O}^\text{t}[\phi]=\int_{\partial\Sigma}\tr(\phi\delta\tilde{\omega}),
\ee
which shows that the generators are integrable. They satisfy the algebra
\be
\lb\widetilde{\O}^\text{t}[\phi],\widetilde{\O}^\text{t}[\varphi]\rb=\Omega_{\partial\Sigma}[e,\omega,u,\chi,\Delta_\phi^\text{t},\Delta_\varphi^\text{t}]=0,
\ee
and are observables in the sense that
\be
\lb\G[\alpha],\widetilde{\O}^\text{t}[\varphi]\rb=\Omega_{\partial\Sigma}[e,\omega,u,\chi,\delta_\alpha^\text{g},\Delta_\varphi^\text{t}]=0,\q\lb\F[\phi],\widetilde{\O}^\text{t}[\varphi]\rb=\Omega_{\partial\Sigma}[e,\omega,u,\chi,\delta_\phi^\text{t},\Delta_\varphi^\text{t}]=0.
\ee
This should be compared with the previous expression \eqref{first order boundary observables 2} for the boundary observable associated to the translations.

Finally, we can compute the Poisson bracket between the observables themselves. This is given by
\be\label{new observables algebra}
\lb\widetilde{\O}^\text{g}[\alpha],\widetilde{\O}^\text{t}[\phi]\rb
&=\Omega_{\partial\Sigma}[e,\omega,u,\chi,\Delta_\alpha^\text{g},\Delta_\phi^\text{t}]\nn\\
&=\int_{\partial\Sigma}\tr\big(\De(u\alpha u^{-1})u\phi u^{-1}\big)\nn\\
&=\widetilde{\O}^\text{t}\big[[\alpha,\phi]\big]+\int_{\partial\Sigma}\tr(\de\alpha\phi),
\ee
which should be put in parallel with \eqref{first order boundary observables algebra}.

Several remarks are now in order. First of all, as announced and as was the case for Abelian Chern--Simons theory, we have obtained a natural disentanglement between the role of gauge transformations and boundary symmetries \cite{Donnelly:2016auv}. The generators \eqref{new SU(2) generator} and \eqref{new translation generator} of gauge transformations are vanishing on-shell and do not require an extension or the restriction to parameters with compact support. Because of the presence of the new boundary fields $u$ and $\chi$, there are two new transformations on the boundary, which are given by \eqref{SU(2) boundary symmetry} and \eqref{topological boundary symmetry}, and which are symmetries since their generators are non-vanishing and constitute observables. At the difference with the observables $\O^\text{g}[\alpha]$ and $\O^\text{t}[\phi]$ obtained previously, the new observables $\widetilde{\O}^\text{g}[\alpha]$ and $\widetilde{\O}^\text{t}[\phi]$ obtained here are defined strongly (i.e. not up to constraints) by boundary integrals, and present a dressing of the connection and triad variables by the new boundary fields. Finally, we can also see that the algebra \eqref{new observables algebra} is strongly equal to a boundary integral, and that this latter reproduces the central term appearing in \eqref{first order boundary observables algebra}.

Finally, before closing this paper we would like to briefly mention the problem of finding a dynamics for the observables and the boundary variables derived above. One possible definition of the boundary dynamics would mean finding a Lagrangian $l[\Phi,\Pi]$ whose variation takes the form
\be
\delta l[\Phi,\Pi]=E[\Phi,\Pi]\wedge\delta(\Phi,\Pi)+\de\vartheta[\Phi,\Pi,\delta\Phi,\delta\Pi],
\ee
where $E[\Phi,\Pi]$ are equations of motion for the original fields $\Phi=(e,\omega)$ and the new boundary fields $\Pi=(u,\chi)$, and where the potential is the corner term appearing in \eqref{fully-invariant first order potential on-shell}, i.e.
\be
\vartheta[\Phi,\Pi,\delta\Phi,\delta\Pi]=\tr\big((e+\De\chi)\delta uu^{-1}-\chi\delta\omega\big).
\ee
This is for example what has been done in \cite{Freidel:2015gpa} (although in another context). Recall first of all, as explained at the end of section \ref{subsubsec:Gauge-invariant boundary-extended Lagrangian}, that the boundary Lagrangian which could be added in order to make \eqref{first order gravity Lagrangian} gauge-invariant is not the appropriate candidate to describe the boundary dynamics along these lines. At first sight, equation \eqref{dressed potential} actually suggests considering $\tr(\tilde{\omega}\wedge\tilde{e})$ as the boundary Lagrangian. Unfortunately, this possibility does not work either because the term $\tr(\tilde{\omega}\wedge\delta\tilde{e})$ leads to unwanted corner contributions which, as one can show, cannot be cancelled even by adding further terms to the Lagrangian. However, it could be that one can bypass this problem by adding further degrees of freedom canonically conjugated to $u$ and $\chi$, but we choose to leave a more detailed analysis of this question for the future.

\section{Conclusion and perspectives}
\label{sec:5}

The general understanding of boundary observables and boundary degrees of freedom in field theory, be it for finite boundaries or boundaries at infinity, is still an open question. Indeed, although there exist many examples in which boundary degrees of freedom and their dynamics can be described, often in connection with very interesting physical applications, there is no systematic derivation or understanding (we believe) of the mechanisms at play. As we have recalled in \ref{sec:2.3}, this depends strongly for example on a choice of boundary conditions or boundary term. What is however clear is that boundary degrees of freedom are tightly connected to the fate of gauge transformations and gauge invariance.

In this article, we have adopted the point of view of \cite{Donnelly:2016auv}, and tried to understand the relationship between the breaking/restauration of gauge-invariance at boundaries and possible associated observables. For this, we have first reviewed in section \ref{sec:2} the covariant Hamiltonian formalism while paying attention to corner ambiguities and boundary terms, and then explained the generic transformation properties of the pre-symplectic potential under finite and field-dependent gauge transformations. We have given in particular a general expression for the extended gauge-invariant potential, which is obtained by introducing boundary fields that restore gauge-invariance at the boundary, and explained how a modification of the usual conserved pre-symplectic form leads to an additional boundary symplectic structure. This is a formalization the idea put forward in \cite{Donnelly:2016auv}. It has the advantage of clearly separating the role of gauge transformations and that of gauge symmetries, the former being generated by constraints and the latter by observables. In summary, the observables arise in this picture as the generators of a new boundary symmetry, which itself arises from the introduction of new boundary fields that ensure gauge-invariance of the potential.

In section \ref{sec:3}, we have applied the general framework of section \ref{sec:2} to the explicit example of Abelian Chern--Simons theory. This constitutes an ideal testbed because it is a theory for which the boundary observables and the boundary dynamics are known. We have found that the observables get dressed by the new boundary field ensuring gauge-invariance, taking the form \eqref{Abelian CS new obervables}, and that their Poisson bracket reproduces that of the affine Kac--Moody algebra \eqref{Abelian CS algebra}. This is in full agreement with the results which have been known for quite some time now, but it makes the whole derivation conceptually clearer. What we have left out of our discussion is the derivation of a dynamics for the boundary observables.

Finally, in section \ref{sec:4} we have studied three-dimensional gravity in its first order formulation. There, not all the gauge symmetries are independent because of the relation \eqref{diffeo gauge and top}, so we have chosen to focus on the $\SU(2)$ gauge transformations and the translations. We have written for the first time the finite form of the translations in the case of a non-vanishing cosmological constant, which is given by \eqref{finite translation e lambda} and \eqref{finite translation omega lambda}. From this, one can however see that the action of finite field-dependent translations on the potential depends strongly on whether the cosmological constant is vanishing or not, and we have chosen the former case for simplicity. After having introduced new boundary fields which ensure the gauge-invariance of the potential under both the $\SU(2)$ gauge transformations and the translations, we have shown that these gauge transformations are generated at the infinitesimal level by generators which vanish on-shell and have a closed (i.e. non-anomalous) algebra. The boundary observables \eqref{new boundary observable SU(2)} and \eqref{new boundary observable top} which we have then obtained are a dressed version of the observables \eqref{first order boundary observables 1} and \eqref{first order boundary observables 2} derived in \cite{Husain:1997fm}, and satisfy the affine Kac--Moody algebra \eqref{first order boundary observables algebra}.

These results show that the extended phase space formalism recovers, via the new boundary fields and dressed boundary observables, the boundary symmetry algebra which was known in the case of Chern--Simons theory and three-dimensional gravity. It opens however the possibility on the one hand of accessing the boundary observables and symmetries in the case of four-dimensional gravity and diffeomorphism transformations (the results of the extended phase space formulation in this case have been presented in \cite{Donnelly:2016auv}), and on the other hand of accessing the boundary action describing the dynamics of the boundary degrees of freedom.

Let us stress here once again that one of the main results of this paper, besides the construction of the gauge-invariant extended symplectic structure for Chern--Simons theory and first order gravity, is the fact that the boundary observables obtained via the extended phase space construction of \cite{Donnelly:2016auv} and which generate the boundary symmetries are a dressed version of the standard Hamiltonian observables \cite{Regge:1974zd,Balachandran:1995qa,Husain:1997fm}. We have found that this dressing corresponds simply to the gauge action of the new boundary fields on the bulk fields incoming at the boundary. This result follows in fact straightforwardly from the way in which the extended phase space has been constructed, which is by acting with gauge transformations on the bulk fields and then promoting the parameters of the gauge transformations to new dynamical degrees of freedom (and thereby automatically introducing the dressing). Beyond the mathematical details of this construction, which has been the scope of this paper, what remains open and challenging to understand is the physical meaning of this dressing and of the new boundary degrees of freedom responsible for it. To understand the physical meaning of these fields, one promising avenue would be to connect the present construction and that of \cite{Donnelly:2016auv}, which were carried out for boundaries at a finite distance, to (possibly null) boundaries at infinity, which is where one describes electromagnetic and gravitational radiation, and where the wealth of recent results summarized in \cite{Strominger:2017zoo} manifest themselves.

Eventually, we are of course interested in understanding the role of these boundary symmetries for quantum gravity. There are already several results in this direction in e.g. loop quantum gravity \cite{Freidel:2015gpa,Freidel:2016bxd}, and this was one of our motivations for studying first order gravity and not second order metric gravity as was done in \cite{Donnelly:2016auv}. In order to go to the four-dimensional case, we will however have to consider both the $\SU(2)$ gauge transformations and the diffeomorphisms, and therefore find the four-dimensional generalization of the extended potential \eqref{extended potential for SU(2) and diff}.

\acknowledgments

I would like to thank Bianca Dittrich, Laurent Freidel, Florian Girelli, Henrique Gomes, Florian Hopfm\"uller, Aldo Riello, Vasudev Shyam, and Wolfgang Wieland for valuable discussions and comments. This research is supported by Perimeter Institute for Theoretical Physics. Research at Perimeter Institute is supported by the Government of Canada through the Department of Innovation, Science and Economic Development, and by the Province of Ontario through the Ministry of Research \& Innovation.

\appendix
\addtocontents{toc}{\protect\setcounter{tocdepth}{1}}

\section{Useful formulas}

For $\mathfrak{g}$-valued differential forms, say $P$ and $Q$ of degree $p$ and $q$, we have denoted in the main text the Lie bracket by
\be
[P\wedge Q]:\Omega^{p+q}(\mathfrak{g}\otimes\mathfrak{g})\rightarrow\Omega^{p+q}(\mathfrak{g}).
\ee
With this, we have the useful permutation and cyclicity relations
\be
[P\wedge Q]=(-1)^{pq+1}[Q\wedge P],\q[P\wedge Q]\wedge R=(-1)^{p(q+r)}[Q\wedge R]\wedge P,
\ee
the Leibniz rule
\be
\De[P\wedge Q]=[\De P\wedge Q]+(-1)^p[P\wedge\De Q],
\ee
the integration by parts formula
\be\label{IBP}
\De P\wedge Q=\de(P\wedge Q)+(-1)^{p+1}P\wedge\De Q,
\ee
and finally the squared action of the covariant derivative is given by
\be
\De\De P=[F\wedge P],
\ee
where $F$ is the curvature.

\section{Pre-symplectic form and generators with boundaries}
\label{appendix:generators}

In this appendix, we start by recalling where the equalities \eqref{gauge generator and 2-form} and \eqref{gauge generator algebra} come from. For this, let us introduce abstract coordinates $\text{A},\text{B},\dots$ on the (infinite-dimensional) phase space. With this, we have the notation
\be
\Omega[\delta_1,\delta_2]=\Omega_\text{AB}(\delta_1\Phi)^\text{A}(\delta_2\Phi)^\text{B},
\ee
where $(\delta\Phi)^\text{A}$ is a tangent vector, and therefore \eqref{gauge generator and 2-form} takes the form
\be
\delta\H[\epsilon]=\Omega_\text{AB}(\delta\Phi)^\text{A}(\delta_\epsilon\Phi)^\text{B}.
\ee
This in turns implies that
\be\label{generator and form relation}
\f{\delta\H[\epsilon]}{(\delta\Phi)^\text{A}}=\Omega_\text{AB}(\delta_\epsilon\Phi)^\text{B}.
\ee
We can now calculate the Poisson bracket between $\H[\epsilon]$ and an arbitrary phase space function $f$, which by definition is given in terms of the inverse pre-symplectic form by
\be
\lb\H[\epsilon],f\rb=\big(\Omega^{-1}\big)^\text{AB}\f{\delta\H[\epsilon]}{(\delta\Phi)^\text{A}}\f{\delta f}{(\delta\Phi)^\text{B}}=-(\delta_\epsilon\Phi)^\text{A}\f{\delta f}{(\delta\Phi)^\text{A}}=\delta_\epsilon f.
\ee
It then follow that
\be
\delta_{\epsilon_1}\H[\epsilon_2]=\lb\H[\epsilon_1],\H[\epsilon_2]\rb=\Omega_\text{AB}(\delta_{\epsilon_1}\Phi)^\text{A}(\delta_{\epsilon_2}\Phi)^\text{B},
\ee
which is \eqref{gauge generator algebra}. Note that we have included a minus sign in our definition of $\delta_\epsilon f$.

Now, imagine that one has an arbitrary variational quantity $\delta\H[\epsilon]$ which has a bulk and a boundary contribution, i.e. which is of the form
\be
\delta\H[\epsilon]=\delta\H_\Sigma[\epsilon]+\delta\H_{\partial\Sigma}[\epsilon].
\ee
If there is no relation of the form \eqref{generator and form relation} between these terms and the pre-symplectic form, it is impossible in general to compute the Poisson bracket between two such phase space functions along these lines. However, as we encounter in the main text, if the pre-symplectic form has a boundary contribution, i.e. if
\be
\Omega[\delta_1,\delta_2]=\Omega_\Sigma[\delta_1,\delta_2]+\Omega_{\partial\Sigma}[\delta_1,\delta_2],
\ee
and if the two variational contributions are defined by
\be
\delta\H_\Sigma[\epsilon]\coloneqq\Omega_\Sigma[\delta,\delta_\epsilon],\q\delta\H_{\partial\Sigma}[\epsilon]\coloneqq\Omega_{\partial\Sigma}[\delta,\delta_\epsilon],
\ee
then one can see that the above calculation goes through, i.e. that it is possible to compute the Poisson bracket
\be
\lb\H_\Sigma[\epsilon_1]+\H_{\partial\Sigma}[\epsilon_1],\H_\Sigma[\epsilon_2]+\H_{\partial\Sigma}[\epsilon_2]\rb=\Omega[\delta_{\epsilon_1},\delta_{\epsilon_2}].
\ee
This explains a subtle yet fundamental difference between the quantities \eqref{variation of abelian CS flatness constraint} and \eqref{Abelian CS generator with symplectic form}.

\section{Details of some calculations}
\label{appendix:proofs}

\paragraph{Proof of \eqref{transformation of original Abelian CS potential a}}
\be
\theta[\alpha^*A,\delta(\alpha^*A)]
&=\delta(A+\de\alpha)\wedge(A+\de\alpha)\nn\\
&=\theta[A,\delta A]+\delta A\wedge\de\alpha+\delta\de\alpha\wedge(A+\de\alpha)\nn\\
&=\theta[A,\delta A]+\delta A\wedge\de\alpha+\de\big(\delta\alpha(A+\de\alpha)\big)-\delta\alpha\de A\nn\\
&=\theta[A,\delta A]+\delta(A\wedge\de\alpha)-A\wedge\delta\de\alpha+\de\big(\delta\alpha(A+\de\alpha)\big)-\delta\alpha\de A\nn\\
&=\theta[A,\delta A]+\delta(A\wedge\de\alpha)-\delta\alpha\de A+\de(\delta\alpha A)+\de\big(\delta\alpha(A+\de\alpha)\big)-\delta\alpha\de A\nn\\
&=\theta[A,\delta A]+\de\big(\delta\alpha(2A+\de\alpha)\big)+\delta(A\wedge\de\alpha)-2\delta\alpha\de A\nn\\
&=\theta[A,\delta A]+\de\big(\delta\alpha(A+\de\alpha)-\alpha\delta A\big)+\delta(\alpha\de A)-2\delta\alpha\de A.
\ee

\paragraph{Proof of \eqref{transformation of non-Abelian CS potential}}~\\
First, we have that
\be
\theta[g^*A,\delta(g^*A)]
&=\la\delta(g^{-1}Ag+g^{-1}\de g)\wedge(g^{-1}Ag+g^{-1}\de g)\ra\nn\\
&=\big\la\big(\delta A+\De(\delta gg^{-1})\big)\wedge(A+\de gg^{-1})\big\ra\nn\\
&=\theta[A,\delta A]+\la\delta A\wedge\de gg^{-1}\ra+\la\De(\delta gg^{-1})\wedge(A+\de gg^{-1})\ra.
\ee
The second term can be rewritten as
\be
\la\delta A\wedge\de gg^{-1}\ra
&=\delta\la A\wedge\de gg^{-1}\ra-\la A\wedge\delta(\de gg^{-1})\ra\nn\\
&=\delta\la A\wedge\de gg^{-1}\ra-\la A\wedge g\de(g^{-1}\delta g)g^{-1}\ra\nn\\
&=\delta\la A\wedge\de gg^{-1}\ra-\la g^{-1}Ag\wedge\de(g^{-1}\delta g)\ra\nn\\
&=\delta\la A\wedge\de gg^{-1}\ra-\la\de(g^{-1}Ag)g^{-1}\delta g\ra+\de\la(g^{-1}Ag)(g^{-1}\delta g)\ra\nn\\
&=\delta\la A\wedge\de gg^{-1}\ra-\la(\de A-[A\wedge\de gg^{-1}])\delta gg^{-1}\ra+\de\la A\delta gg^{-1}\ra\nn\\
&=\delta\la A\wedge\de gg^{-1}\ra-\la\de A\delta gg^{-1}\ra+\la[A\wedge\de gg^{-1}]\delta gg^{-1}\ra+\de\la A\delta gg^{-1}\ra.
\ee
The third term can be rewritten as
\be
\la\De(\delta gg^{-1})\wedge(A+\de gg^{-1})\ra
&=-\la\De(A+\de gg^{-1})\delta gg^{-1}\ra+\de\la(A+\de gg^{-1})\delta gg^{-1}\ra\nn\\
&=-\la\de A\delta gg^{-1}\ra-\la[A\wedge A]\delta gg^{-1}\ra\nn\\
&\pe-\la\de(\de gg^{-1})\delta gg^{-1}\ra-\la[A\wedge\de gg^{-1}]\delta gg^{-1}\ra\nn\\
&\pe+\de\la(A+\de gg^{-1})\delta gg^{-1}\ra\nn\\
&=-\la\de A\delta gg^{-1}\ra-\la[A\wedge A]\delta gg^{-1}\ra\nn\\
&\pe+\la\de(\delta gg^{-1})\wedge\de gg^{-1}\ra-\la[A\wedge\de gg^{-1}]\delta gg^{-1}\ra\nn\\
&\pe+\de\la A\delta gg^{-1}\ra.
\ee
Putting this together, we get that
\be
\theta[g^*A,\delta(g^*A)]=\theta[A,\delta A]+2\de\la A\delta gg^{-1}\ra+\delta\la A\wedge\de gg^{-1}\ra-2\la\delta gg^{-1}F\ra+\la\de(\delta gg^{-1})\wedge\de gg^{-1}\ra.
\ee

\paragraph{Proof of \eqref{infinitesimal topological translation of first order action}}
\be
\delta_\phi^\text{t}L[e,\omega]
&=\tr\left(\delta_\phi^\text{t}e\wedge\left(F+\f{1}{2\ell^2}[e\wedge e]\right)+\delta_\phi^\text{t}\omega\wedge\De e\right)+\de\,\tr(\delta_\phi^\text{t}\omega\wedge e)\nn\\
&=\tr\left(\De\phi\wedge\left(F+\f{1}{2\ell^2}[e\wedge e]\right)+\f{1}{\ell^2}[e,\phi]\wedge\De e\right)+\f{1}{\ell^2}\de\,\tr([e,\phi]\wedge e)\nn\\
&=\tr\left(-\phi\left(\De F+\f{1}{2\ell^2}\De[e\wedge e]\right)+\f{1}{\ell^2}[e,\phi]\wedge\De e\right)
+\de\,\tr\left(\phi\left(F+\f{1}{2\ell^2}[e\wedge e]\right)-\f{1}{\ell^2}\phi[e\wedge e]\right)\nn\\
&=\de\,\tr\left(\phi\left(F+\f{1}{2\ell^2}[e\wedge e]\right)-\f{1}{\ell^2}\phi[e\wedge e]\right).
\ee

\paragraph{Proof of \eqref{finite SU(2) transformation of EOM term}}
\be
E[h^*\Phi]\wedge\delta(h^*\Phi)
&=\tr\left(\delta(h^*e)\wedge\left(h^*F+\f{1}{2\ell^2}[h^*e\wedge h^*e]\right)+\delta(h^*\omega)\wedge h^*(\De e)\right)\nn\\
&=E[\Phi]\wedge\delta\Phi+\tr\left([e,\delta hh^{-1}]\wedge\left(F+\f{1}{2\ell^2}[e\wedge e]\right)+\De(\delta hh^{-1})\wedge\De e\right)\nn\\
&=E[\Phi]\wedge\delta\Phi+\tr\big([e,\delta hh^{-1}]\wedge F+\De(\delta hh^{-1})\wedge\De e\big)\nn\\
&=E[\Phi]\wedge\delta\Phi+\tr\big([e,\delta hh^{-1}]\wedge F+\De e\wedge\De(\delta hh^{-1})\big)\nn\\
&=E[\Phi]\wedge\delta\Phi+\tr\big([e,\delta hh^{-1}]\wedge F+e\wedge\De\De(\delta hh^{-1})\big)+\de\,\tr\big(e\wedge\De(\delta hh^{-1})\big)\nn\\
&=E[\Phi]\wedge\delta\Phi+\tr\big([e,\delta hh^{-1}]\wedge F+e\wedge[F,\delta hh^{-1}]\big)+\de\,\tr\big(e\wedge\De(\delta hh^{-1})\big)\nn\\
&=E[\Phi]\wedge\delta\Phi+\de\,\tr\big(e\wedge\De(\delta hh^{-1})\big).
\ee

\paragraph{Proof of \eqref{finite topological translation transformation of EOM term}}
\be
E[\phi^*\Phi]\wedge\delta(\phi^*\Phi)
&=\tr\big(\delta(\phi^*e)\wedge(\phi^*F)+\delta(\phi^*\omega)\wedge\phi^*(\De e)\big)\nn\\
&=E[\Phi]\wedge\delta\Phi+\tr\big(\delta(\De\phi)\wedge F+\delta\omega\wedge\De\De\phi\big)\nn\\
&=E[\Phi]\wedge\delta\Phi+\tr\big(\delta\de\phi\wedge F+[\delta\omega,\phi]\wedge F+[\omega,\delta\phi]\wedge F+\delta\omega\wedge[F,\phi]\big)\nn\\
&=E[\Phi]\wedge\delta\Phi+\tr\big(-\delta\phi\de F+[\omega,\delta\phi]\wedge F\big)+\de\,\tr(\delta\phi F)\nn\\
&=E[\Phi]\wedge\delta\Phi+\tr(-\delta\phi\De F)+\de\,\tr(\delta\phi F)\nn\\
&=E[\Phi]\wedge\delta\Phi+\de\,\tr(\delta\phi F).
\ee

\section{Non-Abelian Chern--Simons theory}
\label{appendix:CS}

Here we extend partially the results of section \ref{sec:3} to the case of non-Abelian Chern--Simons theory. We will see that the only extra difficulty in doing so is that one has to deal with the WZNW contribution to the potential.

\subsection{Lagrangian}

Let us consider the Lagrangian\footnote{In this paper we always work in the adjoint representation, so that the non-Abelian field strength is given by $F=\de A+[A\wedge A]/2$.}
\be
L[A]\coloneqq\left\la A\wedge\de A+\f{1}{3}A\wedge[A\wedge A]\right\ra=\left\la A\wedge F-\f{1}{6}A\wedge[A\wedge A]\right\ra,
\ee
where $\la\cdot\ra$ denotes a choice of pairing for the Lie algebra elements. Its variation is given by
\be
\delta L[A]=2\la\delta A\wedge F\ra+\de\la\delta A\wedge A\ra,
\ee
from which we can identify the non-Abelian potential. Under infinitesimal and finite gauge transformations, the fields transform as
\be
\delta_\alpha A=\De\alpha,\q\delta_\alpha F=[F,\alpha],\q g^*A=g^{-1}Ag+g^{-1}\de g,\q g^*F=g^{-1}Fg.
\ee
The Lagrangian, on the other hand, transforms as
\be\label{infinitesimal transfo of non-Abelian CS}
\delta_\alpha L[A]=\la\de\alpha\wedge\de A\ra=-\de\la\de\alpha\wedge A\ra=\de\la\alpha\de A\ra,
\ee
and
\be\label{CS finite transformation}
L[g^*A]=L[A]-\f{1}{6}\la\de gg^{-1}\wedge[\de gg^{-1}\wedge\de gg^{-1}]\ra+\de\la A\wedge\de gg^{-1}\ra,
\ee
where the extra term is the sum of a boundary term and the WZNW bulk term.

\subsection{Hamiltonian}
\label{subsec:Hamiltonian CS}

The Hamiltonian action is
\be
S[A]=\int_\mathbb{R}\de t\int_\Sigma\la\partial_0A\wedge A+2A_0F-\de(AA_0)\ra,
\ee
with canonical Poisson bracket $\lb A_a(x),A_b(y)\rb=-\teps_{ab}\delta^2(x,y)/2$, and generic brackets
\be
\lb f_1,f_2\rb=-\f{1}{2}\teps_{ab}\int_\Sigma\de^2x\int_\Sigma\de^2y\,\delta^2(x,y)\f{\delta f_1}{\delta A_a(x)}\f{\delta f_2}{\delta A_b(y)}.
\ee
Let us consider the smeared constraint
\be
\F[\alpha]\coloneqq-2\int_\Sigma\la\alpha F\ra\simeq0,
\ee
whose variation is given by
\be
\delta\F[\alpha]=-2\int_\Sigma\la\delta\alpha F+\delta A\wedge\De\alpha\ra-2\int_{\partial\Sigma}\la\alpha\delta A\ra.
\ee
Now we can proceed as for \eqref{variation of abelian CS flatness constraint}. Let us therefore define an extended generator by
\be
\slashed{\delta}\F_c[\alpha]\coloneqq\delta\F[\alpha]+2\int_{\partial\Sigma}\la\alpha\delta A\ra.
\ee
For $\alpha=\xi\ip A$ we get
\be
\lb\F_c[\xi\ip A],A\rb=\De(\xi\ip A)+\xi\ip F=\de(\xi\ip A)+\xi\ip\de A=\L_\xi A.
\ee
Adding now the condition that $\delta\alpha=0$, we get the generator of gauge transformations
\be
\lb\F_c[\alpha],A\rb=\De\alpha,
\ee
and we can compute the algebra
\be
\lb\F_c[\alpha],\F_c[\beta]\rb=\F_c\big[[\alpha,\beta]\big]+2\int_{\partial\Sigma}\la\de\alpha\beta\ra.
\ee
Note that to obtain this result we have used the identity
\be
\De\alpha\wedge\De\beta=[\alpha,\beta]F-\de([\alpha,\beta]A)+\de\alpha\wedge\de\beta.
\ee
Once again, this algebra closes if we consider generators $\bar{\alpha}$ which are vanishing on $\partial\Sigma$. In this case, $\F_c[\bar{\alpha}]=\F[\bar{\alpha}]\simeq0$ and we have
\be
\lb\F[\bar{\alpha}],\F[\bar{\beta}]\rb=\F\big[[\bar{\alpha},\bar{\beta}]\big].
\ee

For an arbitrary smearing parameter $\alpha$, let us now consider the quantity
\be
\O[\alpha]\coloneqq-2\int_\Sigma\left\la A\wedge\de\alpha+\f{1}{2}\alpha[A\wedge A]\right\ra=\F[\alpha]+2\int_{\partial\Sigma}\la\alpha A\ra\simeq2\int_{\partial\Sigma}\la \alpha A\ra,
\ee
which is not vanishing (i.e. not a flatness constraint) since $\alpha$ does not have to vanish on $\partial\Sigma$. This is an observable since, because $\bar{\alpha}$ has compact support, we have
\be
\lb\O[\alpha],\F[\bar{\alpha}]\rb=\F\big[[\alpha,\bar{\alpha}]\big]+2\int_{\partial\Sigma}\la[\alpha,\bar{\alpha}]A\ra+2\int_{\partial\Sigma}\la\de\alpha\bar{\alpha}\ra\simeq0.
\ee
Now, for $\alpha$ and $\beta$ such that $\alpha\big|_{\partial\Sigma}=\beta\big|_{\partial\Sigma}$, we have $(\alpha-\beta)\big|_{\partial\Sigma}=0$, so
\be
\O[\alpha]-\O[\beta]=\F[\alpha-\beta]\simeq0,
\ee
showing that the observables $\O[\alpha]$ are located on the edge. Finally, these observables satisfy the algebra
\be\label{non-abelian current algebra}
\lb\O[\alpha],\O[\beta]\rb=\O\big[[\alpha,\beta]\big]+2\int_{\partial\Sigma}\la\de\alpha\beta\ra.
\ee
One can see that all these calculations are simply the non-Abelian generalization of that of section \ref{sec:3}.

\subsection{Extended pre-symplectic potential}

Let us show that the property $\delta(L[g^*A])=(\delta L)[g^*A]$ is here indeed satisfied. To compute the variation of \eqref{CS finite transformation} let us first us the invariance of the trace to rewrite $L[g^*A]$ as
\be
L[g^*A]=L[A]-\f{1}{6}\la g^{-1}\de g\wedge[g^{-1}\de g\wedge g^{-1}\de g]\ra+\de\la A\wedge\de gg^{-1}\ra.
\ee
Using $\delta(g^{-1}\de g)=g^{-1}\de(\delta gg^{-1})g$, we then obtain
\be
\delta(L[g^*A])=2\la\delta A\wedge F\ra-\f{1}{2}\la\de(\delta gg^{-1})\wedge[\de gg^{-1}\wedge\de gg^{-1}]\ra+\de\big(\theta[A,\delta A]+\delta\la A\wedge\de gg^{-1}\ra\big),
\ee
with
\be
\theta[A,\delta A]=\la\delta A\wedge A\ra.
\ee
On the other hand, computing the gauge transformation of the variation of the Lagrangian leads to
\be
(\delta L)[g^*A]
&=2\la\delta A\wedge F\ra+2\la\De(\delta gg^{-1})\wedge F\ra+\de\theta[g^*A,\delta(g^*A)]\nn\\
&=2\la\delta A\wedge F\ra+2\de\la\delta gg^{-1}F\ra+\de\theta[g^*A,\delta(g^*A)],
\ee
where we have used the important property
\be
\delta(g^*A)=g^{-1}\big(\delta A+\De(\delta gg^{-1})\big)g,
\ee
together with the integration by parts formula \eqref{IBP} and the Bianchi identity $\De F=0$. Now, as shown in appendix \ref{appendix:proofs}, the potential transforms as
\be\label{transformation of non-Abelian CS potential}
\theta[g^*A,\delta(g^*A)]
&=\theta[A,\delta A]+2\de\la A\delta gg^{-1}\ra+\delta\la A\wedge\de gg^{-1}\ra-2\la\delta gg^{-1}F\ra\nn\\
&\pe+\la\de(\delta gg^{-1})\wedge\de gg^{-1}\ra.
\ee
Using the identity
\be
\de\la\de(\delta gg^{-1})\wedge\de gg^{-1}\ra=-\f{1}{2}\la\de(\delta gg^{-1})\wedge[\de gg^{-1}\wedge\de gg^{-1}]\ra,
\ee
we then get as expected that $\delta(L[g^*A])=(\delta L)[g^*A]$. Also, notice that by using
\be
\la\de(\delta gg^{-1})\wedge\de gg^{-1}\ra=-\f{1}{2}\la[\de gg^{-1}\wedge\de gg^{-1}]\delta gg^{-1}\ra+\de\la\de gg^{-1}\delta gg^{-1}\ra
\ee
the transformed potential can be written as
\be\label{transformation of non-Abelian CS potential bis}
\theta[g^*A,\delta(g^*A)]
&=\theta[A,\delta A]+\de\la(2A+\de gg^{-1})\delta gg^{-1}\ra+\delta\la A\wedge\de gg^{-1}\ra-2\la\delta gg^{-1}F\ra\nn\\
&\pe-\f{1}{2}\la[\de gg^{-1}\wedge\de gg^{-1}]\delta gg^{-1}\ra,
\ee
and it is then transparent that in the Abelian case this reduces to \eqref{transformation of original Abelian CS potential a}.

The first line of the expressions \eqref{transformation of non-Abelian CS potential} and \eqref{transformation of non-Abelian CS potential bis} for the gauge transformation of the potential is indeed of the general form \eqref{transformation of potential}, but one can now see that there is an extra term in the second line which is neither vanishing on-shell nor a total (exterior of field space) derivative. Therefore, although it is possible to define an extended gauge-invariant potential $\theta_{c,m,p}[A,u,\delta A,\delta u]$ by promoting the gauge parameters $g$ to a field $u$ transforming as $g^*u=g^{-1}u$, one has to be careful about what happens to this term in the current. It can however easily be checked that we have
\be
\delta\big(\de(\delta uu^{-1})\wedge\de uu^{-1}\big)=\de\big(\de(\delta uu^{-1})\delta uu^{-1}\big),
\ee
which is a corner contribution. The bulk symplectic structure will therefore be unchanged and in particular conserved. From the gauge-invariant extended potential it is then straightforward to follow the steps of the derivation of section \ref{sec:3} and to compute the boundary observables as well as their algebra. This latter can easily be shown to reproduce \eqref{non-abelian current algebra}.

\section{Gauge-invariant Lagrangians for the translations}
\label{appendix:top}

Just like in the case of Abelian Chern--Simons theory, there are two simple ways of obtaining a Lagrangian which is invariant under the action of the translations (in the case of a vanishing cosmological constant). This is through the introduction of either a boundary or a bulk field.

\subsection{Gauge-invariant boundary-extended Lagrangian}

In order to obtain strict gauge invariance, we can add a dynamical boundary field $\varphi\in\Omega^0\big(\partial M,\su(2)\big)$ which transforms as $\phi^*\varphi=\varphi-\phi$. Indeed, considering the extended Lagrangian
\be
L_\ell[e,\omega,\varphi]\coloneqq L[e,\omega]+\de\,\tr(\varphi F),
\ee
we get that
\be
L_\ell[\phi^*e,\phi^*\omega,\phi^*\varphi]=L_\ell[e,\omega,\varphi].
\ee
Notice also that this preserves the $\SU(2)$ gauge invariance. Indeed, since these gauge transformations act as $h^*\varphi=h^{-1}\varphi h$ on the new field, we get that
\be
L_\ell[h^*e,h^*\omega,h^*\varphi]=L_\ell[e,\omega,\varphi].
\ee
The symplectic potential for the extended Lagrangian is
\be
\theta_\ell[e,\omega,\varphi,\delta\omega,\delta\varphi]=\theta[e,\delta\omega]+\delta\,\tr(\varphi F).
\ee
Looking at gauge transformations, we get that
\be
\theta_\ell[\phi^*e,\phi^*\omega,\phi^*\varphi,\delta(\phi^*\omega),\delta(\phi^*\varphi)]=\theta_\ell[e,\omega,\varphi,\delta\omega,\delta\varphi]-\de\,\tr(\phi\delta\omega)-\tr(\delta\phi F).
\ee
As expected, when comparing this to \eqref{topological translation of bare potential} we see that the term $\delta m[\omega,\phi]=\delta\,\tr(\phi F)$ has been eliminated.

\subsection{Gauge-invariant bulk-extended Lagrangian}

Adding a bulk term with a field $\varphi$ transforming as $h^*\varphi=h^{-1}\varphi h$ and $\phi^*\varphi=\varphi-\phi$, we can consider the Lagrangian
\be
L[e,\omega,\varphi]\coloneqq\tr\big((e+\De\varphi)\wedge F\big),
\ee
which is manifestly strictly gauge-invariant under all the symmetries under consideration. The variation of this Lagrangian is given by
\be
\delta L[e,\omega,\varphi]=\tr(\delta e\wedge F+\delta\omega\wedge\De e)+\de\,\tr\big(\delta\omega\wedge(e+\De\varphi)+\delta\varphi F\big),
\ee
as can be shown by using the Bianchi identity $\De F=0$ together with $\De\De\varphi=[F,\varphi]$. The equations of motion are therefore unchanged. However, the potential is now given by
\be
\theta[e,\omega,\varphi,\delta\omega,\delta\varphi]=\theta[e,\delta\omega]+\tr(\delta\omega\wedge\De\varphi+\delta\varphi F),
\ee
and it transforms as
\be
\theta[\phi^*e,\phi^*\omega,\phi^*\varphi,\delta(\phi^*\omega),\delta(\phi^*\varphi)]=\theta[e,\omega,\varphi,\delta\omega,\delta\varphi]-\tr(\delta\phi F).
\ee
There is therefore no corner term in this transformation.

\section{Diffeomorphisms}

We have chosen in section \ref{sec:4}  to parametrize the total set of gauge transformations of three-dimensional gravity by the $\SU(2)$ ones and the translations. Here we say a brief word about diffeomorphisms. This will illustrate some of the additional difficulties which appear in this case, and will serve as a basis for future work on four-dimensional first order gravity (since there the translations do not exist).

In order to compute the transformation of the potential under a finite diffeomorphism, we have to use the variational formula (3.5) of \cite{Donnelly:2016auv}. For a diffeomorphism $Y:M\rightarrow M$, this is
\be\label{commuting diffeo and delta}
\delta(Y^*f)=Y^*(\delta f+\L_\Y f)=Y^*\big(\delta f+\delta^\text{d}_\Y f\big),
\ee
where the vector field appearing in the Lie derivative is
\be
\Y^a(x)\coloneqq(\delta Y^a\circ Y^{-1})(x).
\ee
This is analogous to the formulas \eqref{commuting delta and SU(2)} and \eqref{commuting delta and top}. With this, we get that
\be
\theta[Y^*e,\delta(Y^*\omega)]=Y^*\big(\theta[e,\delta\omega]+\tr(\L_\Y\omega\wedge e)\big).
\ee
For the sake of simplicity, let us consider the problem of obtaining an extended potential which is only on-shell gauge-invariant under these finite diffeomorphisms. As discussed in section \ref{sec:2} and exemplified with the study of Abelian Chern--Simons theory, this poses no restrictions at all on the derivation of the pre-symplectic two-form, since this latter is defined in any case from its on-shell conservation. Using the infinitesimal equivalence \eqref{diffeo gauge and top}, we can write
\be
\L_\Y\omega=\delta_{\Y\ip\omega}^\text{g}\omega+\delta_{\Y\ip e}^\text{t}\omega+\Y\ip F\simeq\delta_{\Y\ip\omega}^\text{g}\omega=\De(\Y\ip\omega),
\ee
which then leads to
\be
\theta[Y^*e,\delta(Y^*\omega)]
&\simeq Y^*\Big(\theta[e,\delta\omega]+\tr\big(\De(\Y\ip\omega)\wedge e\big)\Big)\nn\\
&=Y^*\Big(\theta[e,\delta\omega]+\de\,\tr\big(e(\Y\ip\omega)\big)-\tr\big((\Y\ip\omega)\De e\big)\Big)\nn\\
&\simeq Y^*\Big(\theta[e,\delta\omega]+\de\,\tr\big(e(\Y\ip\omega)\big)\Big).
\ee

This formula is fundamentally different from the transformation properties \eqref{SU(2) transformation of bare potential} and \eqref{topological translation of bare potential}. This is because the resulting expression depends here on the diffeomorphism $Y$ via $\Y$ and also the global pullback $Y^*$. In particular, one can see that when forming the pre-symplectic $(0,1)$-form \eqref{pre-symplectic 1-form} we get
\be
\Theta_\Sigma[Y^*e,\delta(Y^*\omega)]\simeq\int_{Y(\Sigma)}\theta[e,\delta\omega]+\int_{Y(\partial\Sigma)}\tr\big(e(\Y\ip\omega)\big),
\ee
which therefore has two sources of gauge-non-invariance. In order to define the extended potential, one has to introduce a coordinate system $X$ which defines the surface $\Sigma$ by $\Sigma=X(\sigma)$ and $\partial\Sigma=X(\partial\sigma)$, where $\sigma$ is an open subset of $\mathbb{R}^3$. With this, we can then consider $\Theta_{\Sigma=X(\sigma)}[e,\delta\omega]$, and since under a diffeomorphism $Y$ we have $Y^*X=Y^{-1}\circ X$, this becomes
\be
\Theta_{\Sigma=X(\sigma)}[Y^*e,\delta(Y^*\omega)]
&\simeq\int_{(Y^{-1}\circ X)(\sigma)}Y^*\Big(\theta[e,\delta\omega]+\de\,\tr\big(e(\Y\ip\omega)\big)\Big)\nn\\
&\simeq\int_{\Sigma=X(\sigma)}\Big(\theta[e,\delta\omega]+\de\,\tr\big(e(\Y\ip\omega)\big)\Big).
\ee
Now, for $\X\coloneqq\delta X\circ X^{-1}$ the analogue for diffeomorphisms of formula \eqref{composition of h and u} is given by
\be
\delta(Y^*X)\circ(Y^*X)^{-1}=\delta(Y^{-1}\circ X)\circ(Y^{-1}\circ X)^{-1}=Y^*(\X-\Y),
\ee
which implies that one can construct an on-shell gauge-invariant extended potential $(0,1)$-form as
\be\label{diffeos extended potential}
\Theta_{\Sigma,\partial\Sigma}[e,X,\delta\omega,\delta X]\coloneqq\int_{\Sigma=X(\sigma)}\theta[e,\delta\omega]+\int_{\partial\Sigma=X(\partial\sigma)}\tr\big(e(\X\ip\omega)\big).
\ee
One can check explicitly that this satisfies $\Theta[Y^*e,Y^*X,\delta(Y^*\omega),\delta(Y^*X)]=\Theta[e,X,\delta\omega,\delta X]$. When constructing the pre-symplectic form from this expression, one then has to be extra careful with the subtlety mentioned below \eqref{pre-symplectic 1-form}, which has to do with the fact that the locations $\Sigma$ and $\partial\Sigma$ are now specified in a field-dependent way. This is described in details in \cite{Donnelly:2016auv} for metric gravity, and we will develop our construction for first order gravity in a subsequent work \cite{MGtoappear}.

As we have already encountered in the main text, the integrability condition for the boundary observables requires (at least if no boundary conditions are put on the fields themselves) that the parameters of the boundary symmetry be field-independent. This means that, starting from the boundary observables corresponding to $\SU(2)$ gauge transformations and translations, one cannot simply use the relations \eqref{relation d t g for e} and \eqref{relation d t g for omega} in order to construct the boundary observables associated with diffeomorphisms. As mentioned above, the study of diffeomorphisms therefore requires a separate study and will appear in a subsequent work.

Now, the same question as in section \ref{subsec:compatibility} arises for the compatibility between $\SU(2)$ gauge transformations and diffeomorphisms. Let us once again go on-shell, since this is all that matters for the computation of the pre-symplectic form. The problem of compatibility has to do with the fact that $\SU(2)$ gauge transformations and diffeomorphisms do not commute. One could therefore think of solving this issue by using the gauge-covariant Lie derivative\footnote{This object has appeared previously and been used in \cite{Jacobson:2015uqa,Freidel:2015gpa} in the context of first order gravity.} defined as $\gL_\xi\coloneqq\L_\xi-\delta^\text{g}_{\xi\ip\omega}$. Its action on the frame field is given by $\gL_\xi e=\De(\xi\ip e)+\xi\ip(\De e)$, which is indeed a covariant version of \eqref{relation d t g for e}, while its action on the connection is given by $\gL_\xi\omega=\xi\ip F$. However, one cannot naively substitute $\gL_\xi$ for $\L_\xi$ in the above formulas, since there would be no interpretation for \eqref{commuting diffeo and delta}. Instead, one can compute explicitly the action of diffeomorphisms and $\SU(2)$ gauge transformations, and then find the additional corner term that makes them commutative, and therefore defines a potential which is fully gauge-invariant.

When acting on a scalar function such as $u$, the Lie derivative is simply the directional derivative $\L_\xi u=\xi\ip\de u$. With this, the on-shell corner term in the potential \eqref{first order SU(2) extended potential} transforms under diffeomorphisms as
\be
\Theta_{\partial\Sigma}[Y^*e,Y^*u,\delta(Y^*u)]=\int_{\partial\Sigma=X(\partial\sigma)}\tr(e\delta uu^{-1})+\int_{\partial\Sigma=X(\partial\sigma)}\tr\big(e(\Y\ip\de uu^{-1})\big).
\ee
On the other hand, if the $\SU(2)$ gauge transformations do not act on $X$, their action on the corner term in the potential \eqref{diffeos extended potential} is given by
\be
\Theta_{\partial\Sigma}[h^*e,h^*X,\delta(h^*X)]=\int_{\partial\Sigma=X(\partial\sigma)}\tr\big(e(\X\ip\omega)\big)+\int_{\partial\Sigma=X(\partial\sigma)}\tr\big(e(\X\ip\de hh^{-1})\big).
\ee
From this, one can see that full on-shell gauge-invariance can be obtained by considering
\be\label{extended potential for SU(2) and diff}
\Theta_{\Sigma,\partial\Sigma}[e,\omega,u,X,\delta\omega,\delta u,\delta X]\coloneqq\int_{\Sigma=X(\sigma)}\theta[e,\delta\omega]+\int_{\partial\Sigma=X(\partial\sigma)}\tr\big(e\delta uu^{-1}+e(\X\ip\omega)+e(\X\ip\de uu^{-1})\big).
\ee
The study of this extended potential and of the associated boundary symmetries will be presented in \cite{MGtoappear}.

\bibliography{Biblio.bib}
\bibliographystyle{JHEP}

\end{document}